\documentclass[12pt,preprint]{aastex}
\usepackage{epstopdf}

\newcommand{\dif}{\mathrm{d}}

\usepackage{amsmath}

\begin{document}

\title{\large \sc Properties of the H$\alpha$-emitting Circumstellar
Regions of \\Be Stars}

\author{Christopher Tycner,\altaffilmark{1,2,3}
John B. Lester,\altaffilmark{4}
Arsen R. Hajian,\altaffilmark{5}
J.~T.~Armstrong,\altaffilmark{6}
J.~A.~Benson,\altaffilmark{1}
G.~C.~Gilbreath,\altaffilmark{6}
D.~J.~Hutter,\altaffilmark{1}
T.~A.~Pauls,\altaffilmark{6}
N.~M.~White\altaffilmark{7}\\
\bigskip
{\tt \footnotesize tycner@nofs.navy.mil; lester@astro.utoronto.ca;
hajian@usno.navy.mil; tom.armstrong@nrl.navy.mil;
jbenson@nofs.navy.mil; Charmaine.Gilbreath@nrl.navy.mil;
djh@nofs.navy.mil; pauls@nrl.navy.mil; nmw@lowell.edu}\\
\bigskip
\bigskip
\bigskip \bigskip
}

\altaffiltext{1}{US Naval Observatory, Flagstaff Station, P.O. Box
1149, Flagstaff, AZ 86002-1149 } \altaffiltext{2}{NVI, Inc., 7257
Hanover Parkway, Suite D, Greenbelt, MD 2077011} \altaffiltext{3}{Also
Department of Astronomy and Astrophysics, University of Toronto,
Toronto, ON, Canada} \altaffiltext{4}{Department of Astronomy and
Astrophysics, Erindale Campus, University of Toronto, 3359 Mississauga
Road North, Mississauga, ON L5L 1C6, Canada} \altaffiltext{5}{US Naval
Observatory, 3450 Massachusetts Avenue, NW, Washington, DC 20392-5420}
\altaffiltext{6}{Remote Sensing Division, Code 7210, Naval Research
Laboratory, 4555 Overlook Avenue, SW, Washington, DC 20375}
\altaffiltext{7}{Lowell Observatory, 1400 West Mars Hill Road,
Flagstaff, AZ 86001}

\begin{abstract}
Long-baseline interferometric observations obtained with the Navy
Prototype Optical Interferometer of the H$\alpha$-emitting envelopes
of the Be stars $\eta$~Tauri and $\beta$~Canis~Minoris are presented.
For compatibility with the previously published interferometric
results in the literature of other Be stars, circularly symmetric and
elliptical Gaussian models were fitted to the calibrated H$\alpha$
observations.  The models are sufficient in characterizing the angular
distribution of the H$\alpha$-emitting circumstellar material
associated with these Be stars.  To study the correlations between the
various model parameters and the stellar properties, the model
parameters for $\eta$~Tau and $\beta$~CMi were combined with data for
other Be stars from the literature.  After accounting for the
different distances to the sources and stellar continuum flux levels,
it was possible to study the relationship between the net H$\alpha$
emission and the physical extent of the H$\alpha$-emitting
circumstellar region.  A clear dependence of the net H$\alpha$
emission on the linear size of the emitting region is demonstrated and
these results are consistent with an optically thick line emission
that is directly proportional to the effective area of the emitting
disk.  Within the small sample of stars considered in this analysis,
no clear dependence on the spectral type or stellar rotation is found,
although the results do suggest that hotter stars might have more
extended H$\alpha$-emitting regions.
\end{abstract}


\section{Introduction}

Over the years, numerous observational and theoretical studies have
shown that nearly all stars lose mass in the form of stellar winds.
Although the details associated with the wind formation can be quite
different from one star to another, and could involve a number of
different physical processes, the most common driving force is the
stellar radiation field.  In fact, the radiation alone can be
responsible for driving the stellar wind as the photons are absorbed
or scattered by the outflowing material.  In cool stars, where the
conditions in the wind permit dust to form, the photons are absorbed
by the dust particles, while in hotter stars the photon momentum is
transferred to the wind by excitation of the atomic gas, resulting in
what is known as line-driven wind.

Among the B-type stars, which have effective temperatures in the range
of 10,000 to 30,000~K, there exists a very interesting class of stars
that possess circumstellar regions with much higher densities than
expected for line-driven winds of stars of similar effective
temperatures and masses.  These stars are known as Be stars and their
common characteristic is that they have, or had in the past, at least
one Balmer line in emission.  Typically, the Be notation is reserved
for non-supergiant B-type stars with line-emitting circumstellar
material.  It is sometimes necessary to emphasize the difference
between Be stars with circumstellar envelopes that are thought to be
formed by an outflowing material and B-type stars with circumstellar
regions that are formed by accretion of the surrounding material, as
in the case of Herbig~Ae/Be stars, or semi-detached~(Algol type)
binary systems.  In such case, the term ``classical Be star'' is used
to explicitly identify non-accreting B-type stars with circumstellar
regions~\citep[see also the review paper by][]{Porter03}.

The initial attempts to apply optical interferometry to Be stars have
been quite promising.  \citet{Thom86} were the first to spatially
resolve the the circumstellar envelope of a Be star ($\gamma$~Cas)
using the I2T interferometer.  Shortly thereafter, the Mark~III
interferometer was used to obtain the first direct measurements of the
deviation from spherical symmetry of the H$\alpha$-emitting regions of
$\gamma$~Cas and $\zeta$~Tau~\citep{Quirrenbach93,Quirrenbach94}.
Interferometric observations that divide the H$\alpha$ emission line
into a number of spectral channels have also been obtained with the
GI2T interferometer, and they resulted in the first detection of a
rotational signature in an envelope of a Be
star~\citep{Mourard89,Stee95}.  The combination of linear polarization
studies with interferometric observations of a number of Be stars,
which have shown that the H$\alpha$-emitting envelopes are flattened,
is the strongest observational evidence to date for the presence of
relatively thin circumstellar disks~\citep{Quirrenbach97}.

The recent interferometric detection of a high degree of oblateness of
the stellar photosphere in the Be star $\alpha$~Eridani
by~\citet{Domiciano03}, compounded with the possibility that most
$v\sin i$ values for Be stars could be significantly
underestimated~\citep{Townsend04}, has reignited the interest in the
connection between the stellar rotation and the Be disk formation
mechanisms.  Similarly, the study of the correlation between the
strength of the H$\alpha$ emission line and the spatial extent of the
emitting region of a Be star can be used to establish the properties
of the circumstellar region.  Although \citet{Quirrenbach97} have
shown that there is a tendency for stars with larger disks to have
stronger H$\alpha$ emission, there was also a considerable scatter
present in their results.  We will extend their analysis to account
for the different base flux levels in the H$\alpha$ emission lines,
and we will show that there is a clear relationship between the net
emission and the physical extent of the H$\alpha$-emitting disk.

\section{Initial Reductions and Calibration}

The new observations that we present in this paper, as well as those
that were reported previously~\citep{Tycner03,Tycner04}, were all
obtained with the Navy Prototype Optical Interferometer~(NPOI).  The
instrument is a descendant of the Mark~I through Mark~III
phase-tracking optical interferometers~\citep{Shao80,Shao87,Shao88}
and is located at the dark observing site of the Lowell Observatory,
near Flagstaff, Arizona.  We will review here only those aspects of
the instrument and the observational setup that are directly relevant
to the work presented in this paper.  For a more detailed discussion
of the instrument, the initial data reduction process, and the
calibration, the reader is referred to the papers
by~\citet{Armstrong98}, \citet{Hummel98}, and \citet{Tycner04},
respectively.

\begin{figure}
\plotone{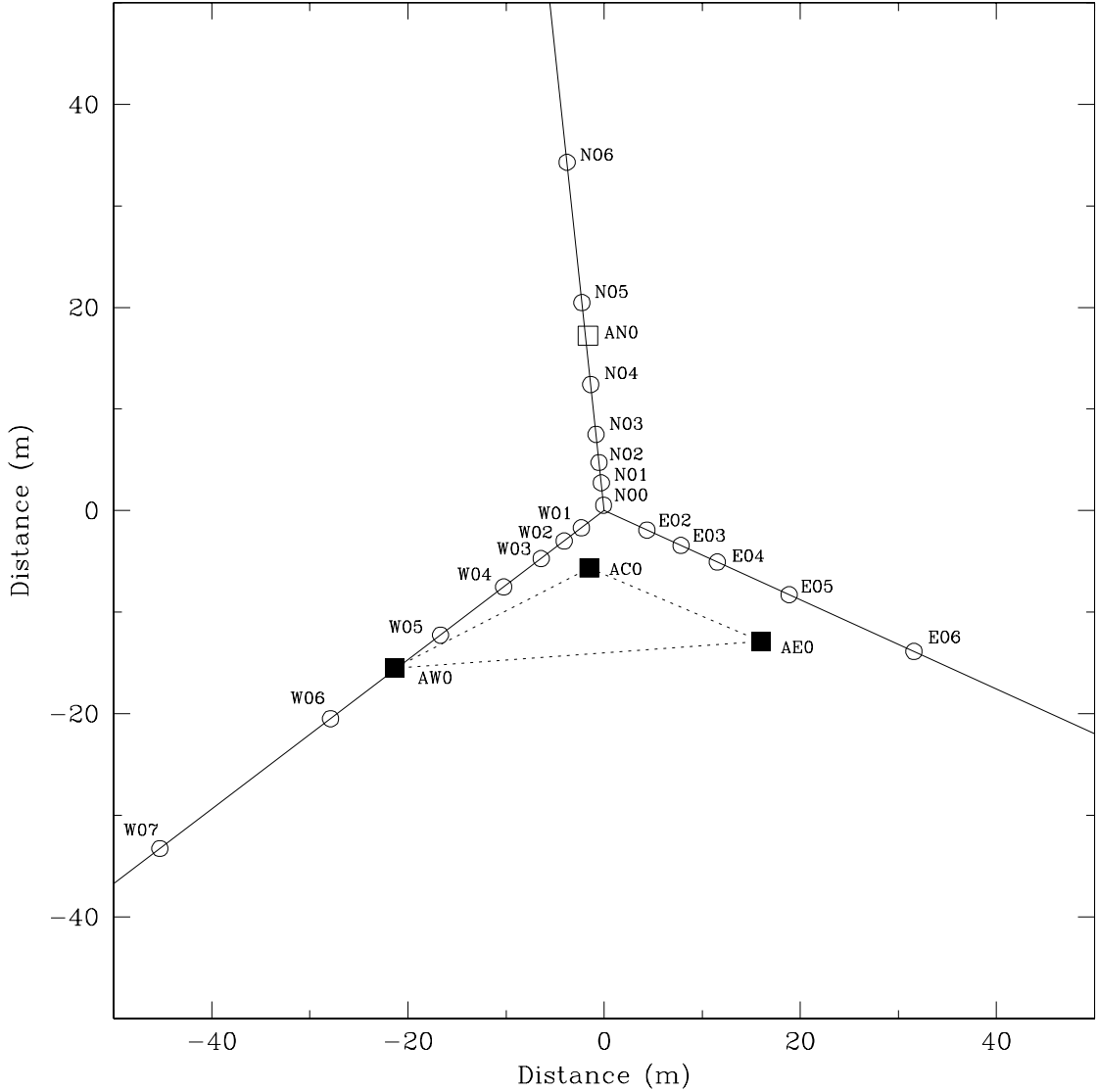}
\caption{A schematic of the inner region of the NPOI array.  Both the
imaging~({\it circles}) and the astrometric~({\it squares}) stations
are shown.  The three astrometric stations used in the observations
presented in this paper are shown with filled squares.  The resulting
baselines~({\it dotted lines}) have lengths of 18.9~(AC0--AE0),
22.2~(AW0-AC0), and 37.5~m~(AW0--AE0) and are at azimuths of
$-67\fdg5$, $63\fdg6$, and $86\fdg0$, respectively.}
\label{fig:NPOI_model_sub}
\end{figure}

The NPOI consists of an imaging and an astrometric array, but for all
the NPOI observations discussed in this paper only three astrometric
stations located in the inner part of the array were used~\citep[also
referred to as the AC0, AE0, and AW0 stations;][]{Armstrong98}.  The
three baselines formed by the AC0--AE0--AW0 triangle are shown in
Figure~\ref{fig:NPOI_model_sub} with dotted lines.  Each pair of
elements creates an interference pattern at the beam combiner, which
is then dispersed by a prism onto a lenslet array.  A series of
optical fibers that are glued to the back of each of the three lenslet
arrays carry the signal to a cluster of photon-counting avalanche
photodiodes~(APDs), which record the intensity variations as a
function of a variable optical path difference.  Each lenslet array
provides 32~spectral channels covering the 450--850~nm spectral
region.  However, because the number of available APDs was limited at
the time of observations, as well as the signal-to-noise ratio~(SNR)
was too low in some channels~(especially those at shorter wavelengths
where the sensitivity of the instrument is low), not all spectral
channels were usable.  Also, the spectral alignment of the lenslet
arrays was such that the H$\alpha$ emission was located in one of two
spectral channels depending on the baseline~(i.e., depending which
lenslet array detected the signal for that baseline).  These two
channels are labeled with central wavelengths of 648 and 665~nm, but
because both of these channels have approximately the same spectral
width of 16~nm, the ones that contain the H$\alpha$ emission line are
referred to as the H$\alpha$ channels~\citep[see also][where the
alignment of the spectral channels is described in detail]{Tycner04}.

We obtain an estimate of squared visibility for every 2~ms data frame,
and these values are then incoherently averaged over 1~s intervals to
create 1~s data points~\citep[see][for more details]{Hummel98}.  In a
subsequent step a raw squared visibility that corresponds to a 90~s
interval~(known also as a ``scan'') is calculated by a simple average.
In the process, the observational errors of the raw scans are set
equal to the standard deviation of the 1~s data points divided by the
square root of the number of points in a scan.  This approach produces
larger errors than one would get by propagating the errors of the
individual 1~s data points~\citep{Hummel98}, but it is part of the
standard reduction scheme that attempts to account for both the white
and the non-white noise sources~(i.e., it attempts to account for some
of the atmospheric and instrumental effects).  If some of these
effects are removed from the data through subsequent
calibration~\citep[like the one we developed in][]{Tycner03}, these
error bars will tend to overestimate the magnitude of the actual
observational uncertainties.  New methods for estimating the
interferometric observables along with their associated uncertainties
are currently being developed by the NPOI team, but their discussion
is beyond the scope of this paper.

\subsection{Observations of $\eta$~Tauri and $\beta$~Canis~Minoris}
\label{sec:eta_tau_obs}

The Be stars $\eta$~Tauri~(=HR~1165) and
$\beta$~Canis~Minoris~(=HR~2845) have been observed with the NPOI on
many occasions, although most of the time not as primary scientific
targets.  In many instances they have been used as calibration
sources, or check stars for observations of other targets, frequently
binary systems.  This resulted in a quite limited coverage of these
stars on any single night, but searching through the archival
observations of NPOI we have found 22 nights from late 1997 to early
1999 that contained at least some valid scans of $\eta$~Tau or
$\beta$~CMi.  Some scans were excluded because of missing data on one
of the baselines, or very low SNR due to poor atmospheric
conditions.  Table~\ref{tab:new_observations} lists the dates and the
number of scans of $\eta$~Tau and $\beta$~CMi available for analysis
from each of the 22 nights.

We begin by calibrating the $V^2$-values of $\eta$~Tau and $\beta$~CMi
using exactly the same method as was used in the case of
$\zeta$~Tau~\citep{Tycner04}.  For the stellar disk diameters of
$\eta$~Tau and $\beta$~CMi we adopt the photometrically inferred
values of 0.71 and 0.74, respectively~\citep{Quirrenbach97}.  Because
the instrumental configuration of the NPOI was unchanged, for all
practical purposes, from 1997 until mid-1999, the spectral alignment
of the output beams with respect to the lenslet arrays was the same
for all stars observed with the NPOI over that period.  Also, as it
was discussed in \citet{Tycner04}, the alignment of the spectral
channels was such that the H$\alpha$ emission was located in either
the 648 or 665~nm channel, depending on the baseline.  We verify if
this is also the case for $\eta$~Tau and $\beta$~CMi by comparing the
signal from the channels that are expected to contain the H$\alpha$
signal with those that do not.

The calibrated squared visibilities of $\eta$~Tau from the
H$\alpha$-containing channels are shown in
Figure~\ref{fig:model_fit-ETau} along with the $V^2$-values for the
uniform disk~(UD) model representing the stellar photosphere.  The
squared visibilities from the H$\alpha$ channels systematically fall
below the expected photospheric values, especially at the higher
spatial frequencies.  We interpret this as a direct signature of a
resolved circumstellar envelope that is responsible for the H$\alpha$
emission.  Figure~\ref{fig:noHalpha-ETau} demonstrates that the
signature in the 648 and 665~nm channels without the H$\alpha$
emission, is very similar to that seen in the continuum channels at
both shorter~(Fig.~\ref{fig:604nm_res-ETau}), and
longer~(Fig.~\ref{fig:702nm_res-ETau}) wavelengths.  Although the
quantity of data is much smaller in the case of $\beta$~CMi, the same
conclusions can be reached by inspecting the signal in the spectral
channels with and without the H$\alpha$ emission~(see
Figs.~\ref{fig:model_fit-BCMi} and~\ref{fig:noHalpha-BCMi}).  Based on
this we conclude that all of the H$\alpha$ emission is captured in one
spectral channel at all three baselines for both $\eta$~Tau and
$\beta$~CMi.

\subsection{Modeling the H$\alpha$-emitting Envelopes}
\label{sec:models}

The interferometric observations of Be stars represent a unique class
of observations, where the star can be modeled with a simple
two-component model.  At the angular resolution provided by our
observations the stellar photosphere, which is visible at the
continuum wavelengths, can be modeled with the UD model.  The second
component, which is responsible for the H$\alpha$ emission, can be
modeled with an elliptical Gaussian to determine the angular extent
and the ellipticity of the circumstellar region.  If the two
components are assumed to be concentric, then the net squared
visibility for both components can be expressed as a normalized sum of
the form:
\begin{equation}
\label{eqn:UDandGauss}
V^2_{\rm UD+Gauss} = \left[ c_{\rm p} V_{\rm UD}(\theta_{\rm UD}) +
(1-c_{\rm p}) V_{\rm Gauss}(\theta_{\rm mj},r,\phi) \right]^2,
\end{equation}
where both $V_{\rm UD}(\theta_{\rm UD})$ and $V_{\rm
Gauss}(\theta_{\rm mj},r,\phi)$ are discussed in detail in
\citet{Tycner04}.  The $c_{\mathrm{p}}$ parameter in the above
equation represents the net fractional contribution from the stellar
photosphere to the net signal in the H$\alpha$ channel, and, because
it is defined to have a value between 0 and 1, it also maintains the
normalization of the net visibility.  Of course, if the two components
are not concentric then the individual visibility amplitudes cannot be
added in such a simple way and one would have to account for the
relative complex phase shifts between the models.  However, because
our observations were obtained at a relatively short
baselines~(18.9--37.5~m), we resolve only the overall shape of the
circumstellar regions and we are insensitive to small scale deviations
from point-symmetric source structure.

\begin{figure}
\plotone{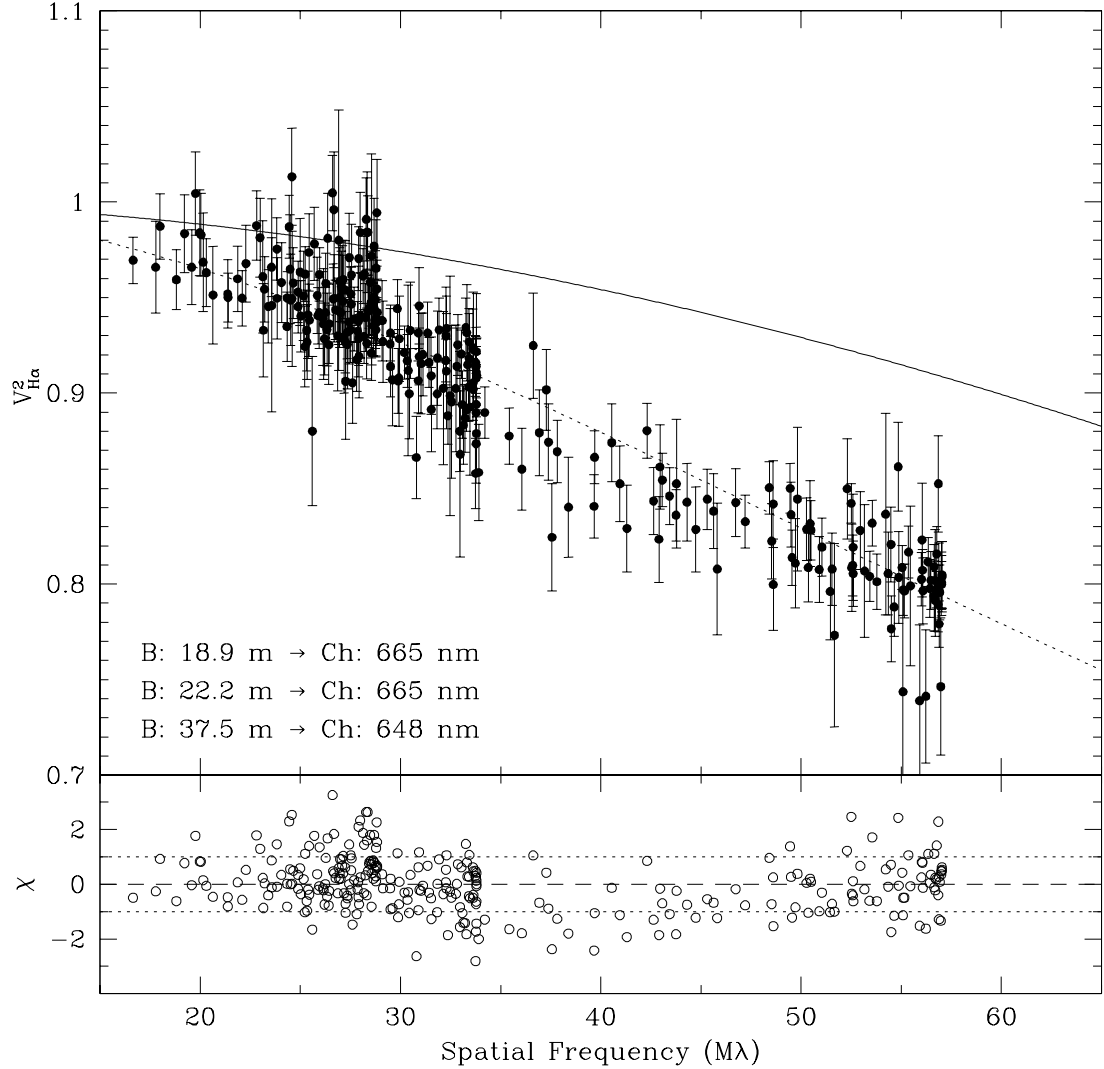}
\caption{Calibrated squared visibilities of $\eta$~Tau from the
spectral channels that contain the H$\alpha$ emission~(i.e., 665~nm
channel for the 18.9 and 22.2~m baselines, and 648~nm channel for the
37.5~m baseline).  The uniform disk model representing the stellar
photospheric disk~({\it solid-line}) and the best-fit circularly
symmetric Gaussian model~({\it dotted-line}) are also shown.  The
normalized residuals for the circularly symmetric Gaussian are shown
in the lower panel. }
\label{fig:model_fit-ETau}
\end{figure}

\begin{figure}
\plotone{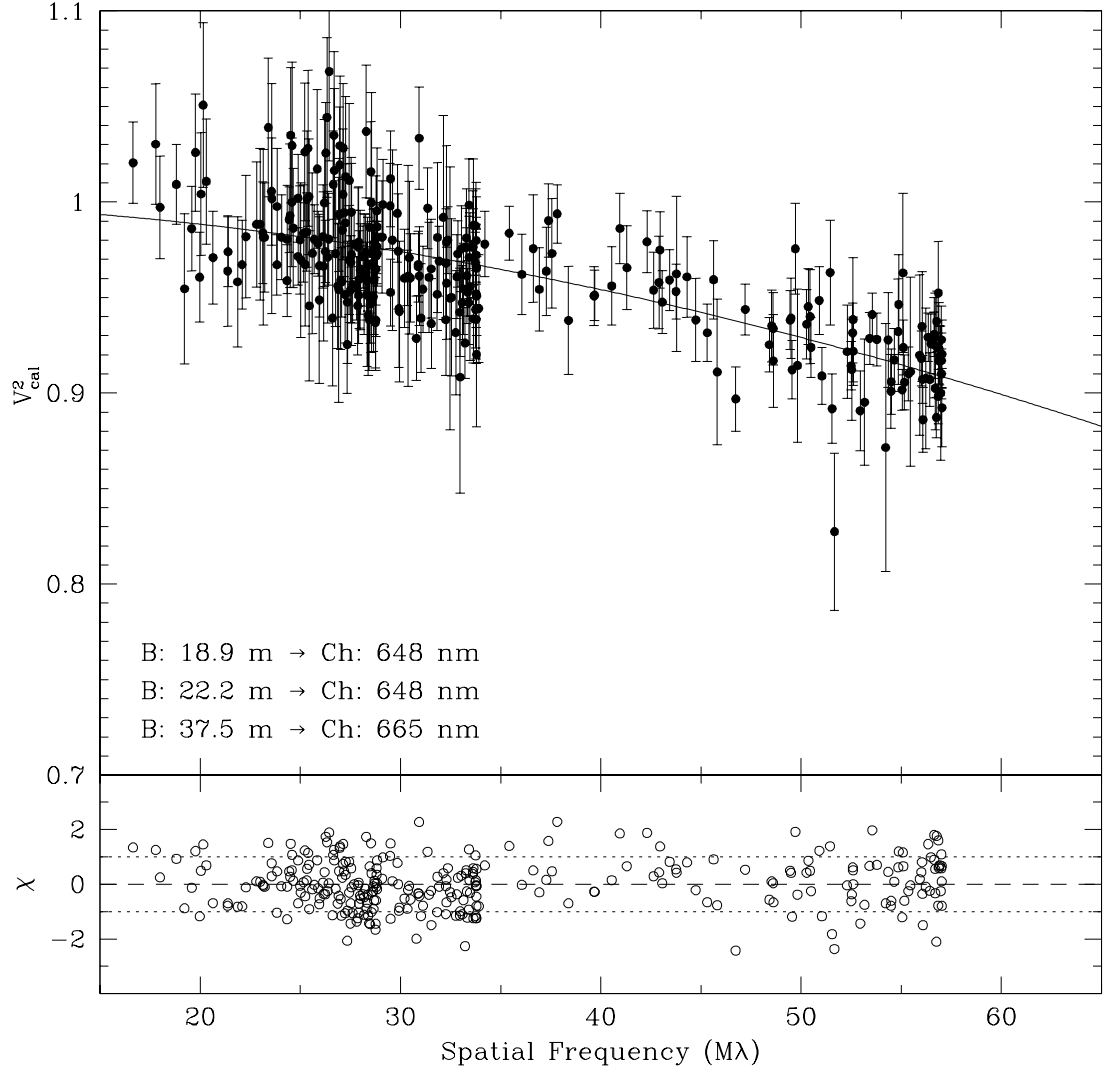}
\caption{Same as Fig.~\ref{fig:model_fit-ETau} but with the choice of
the spectral channels reversed~(i.e., the 648~nm channel was used for
the two shortest baselines and the 665~nm channel for the longest
baseline).  The residuals in the lower panel are calculated with
respect to the uniform disk model representing the stellar
photosphere.  }
\label{fig:noHalpha-ETau}
\end{figure}

\begin{figure}
\plotone{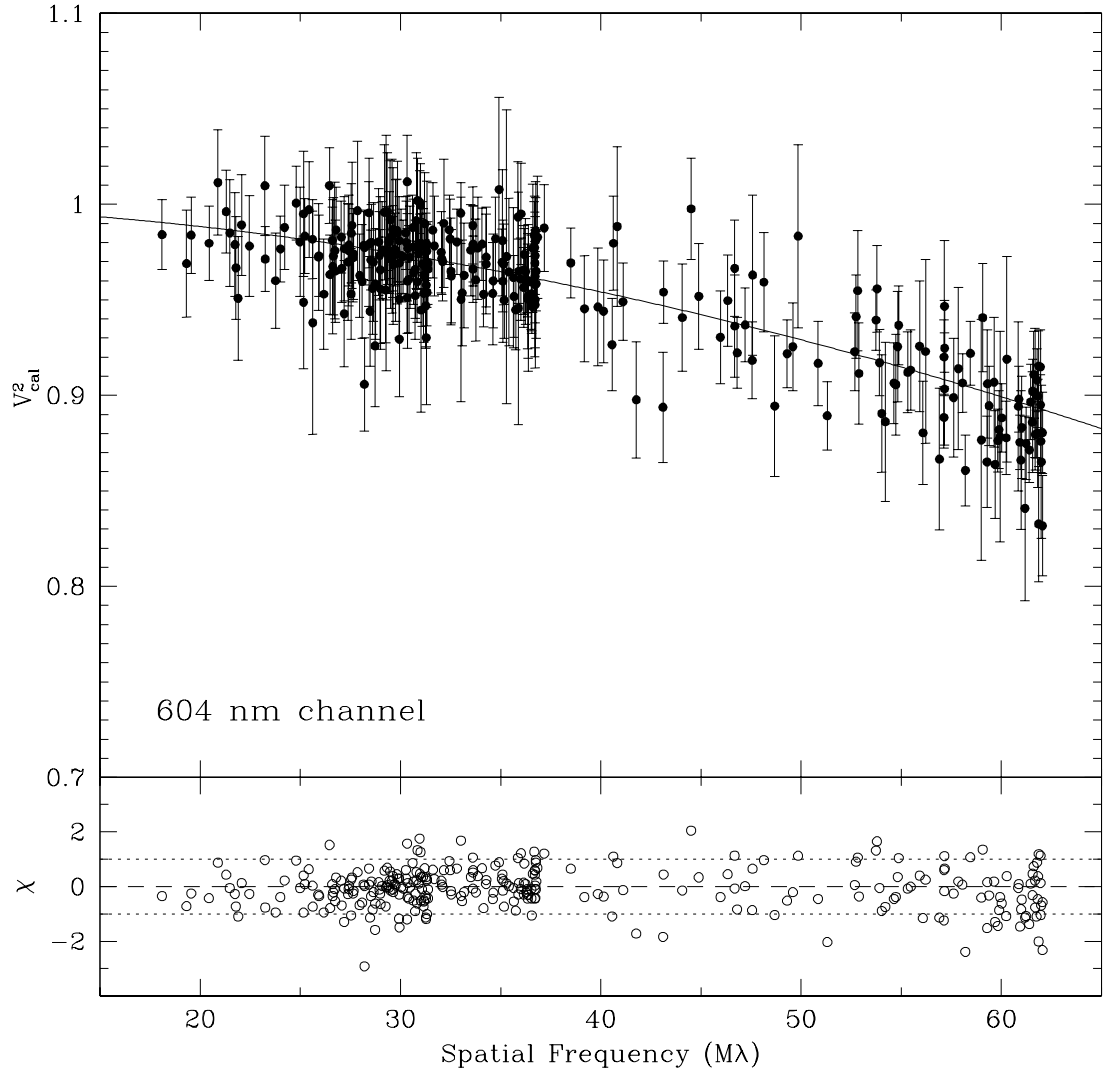}
\caption{Calibrated squared visibilities of $\eta$~Tau from the
continuum channel at 604~nm.  }
\label{fig:604nm_res-ETau}
\end{figure}

\begin{figure}
\plotone{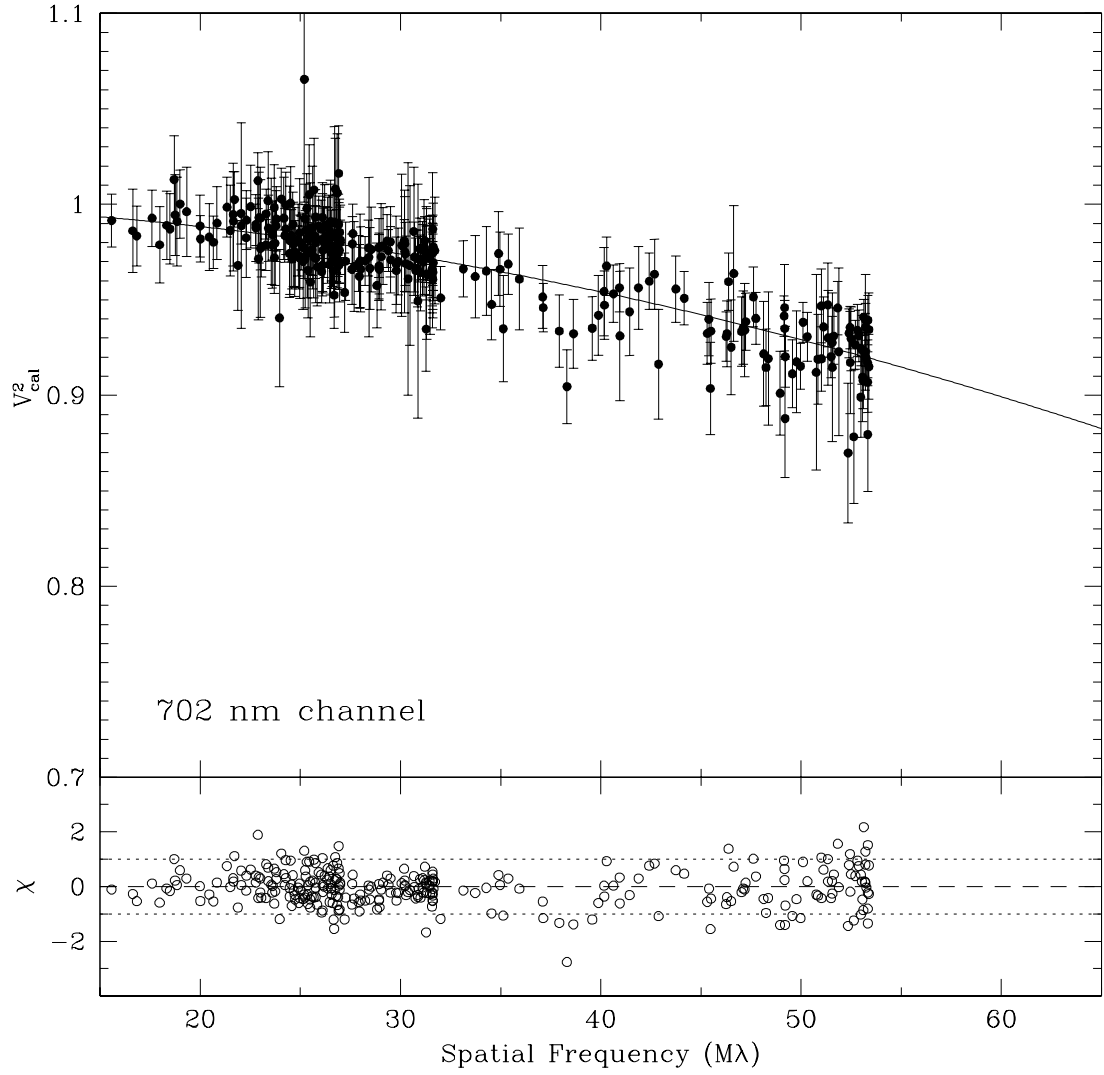}
\caption{Same as Fig.~\ref{fig:604nm_res-ETau} but for the continuum
channel at 702~nm.  }
\label{fig:702nm_res-ETau}
\end{figure}

\begin{figure}
\plotone{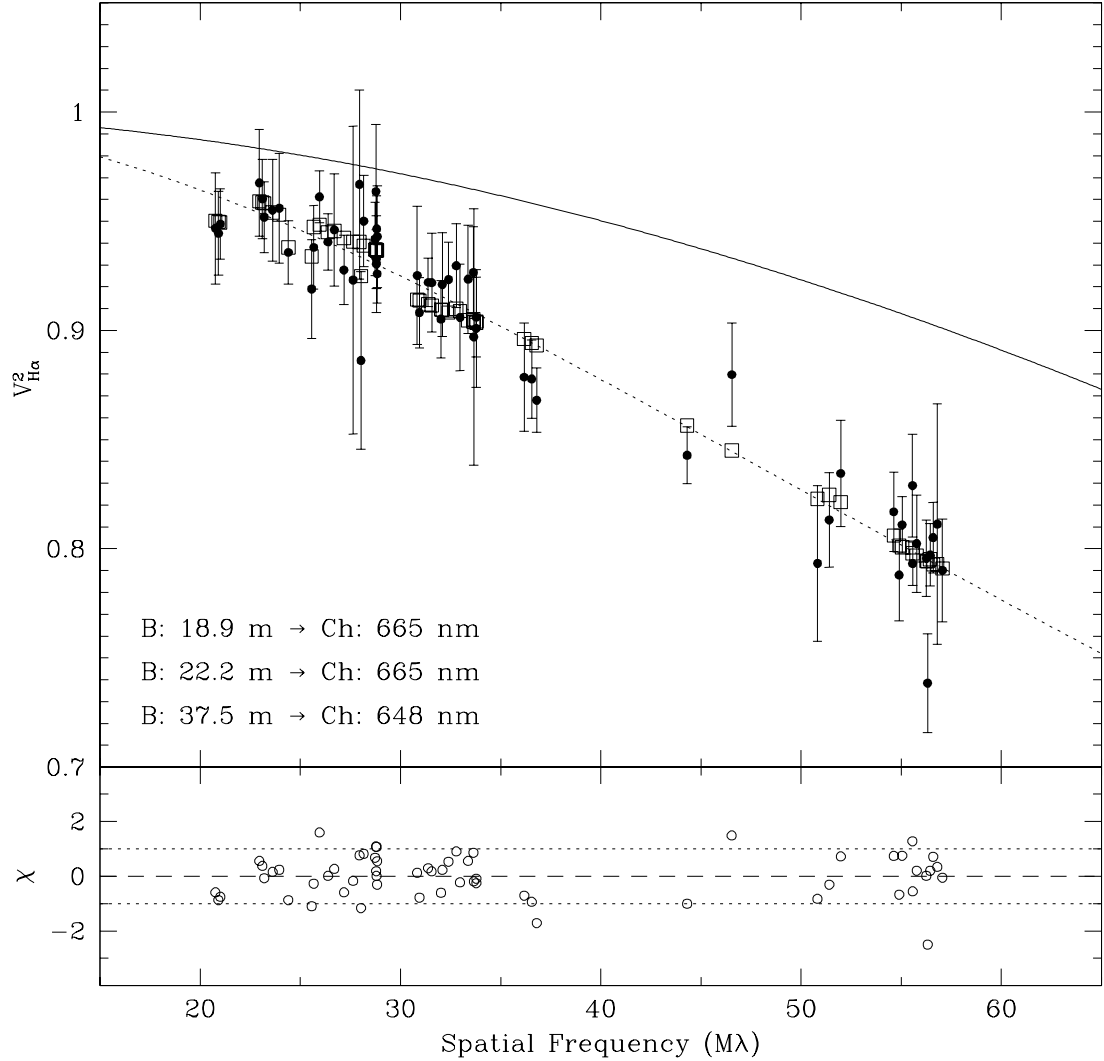}
\caption{Calibrated squared visibilities of $\beta$~CMi from the
spectral channels that contain the H$\alpha$ emission~(i.e., 665~nm
channel for the 18.9 and 22.2~m baselines, and 648~nm channel for the
37.5~m baseline).  The uniform disk model representing the stellar
photospheric disk~({\it solid line}) and the best-fit circularly
symmetric~({\it dotted line}) and elliptical~({\it squares}) Gaussian
models are also shown.  The normalized residuals for the circularly
symmetric Gaussian model are shown in the lower panel. }
\label{fig:model_fit-BCMi}
\end{figure}

\begin{figure}
\plotone{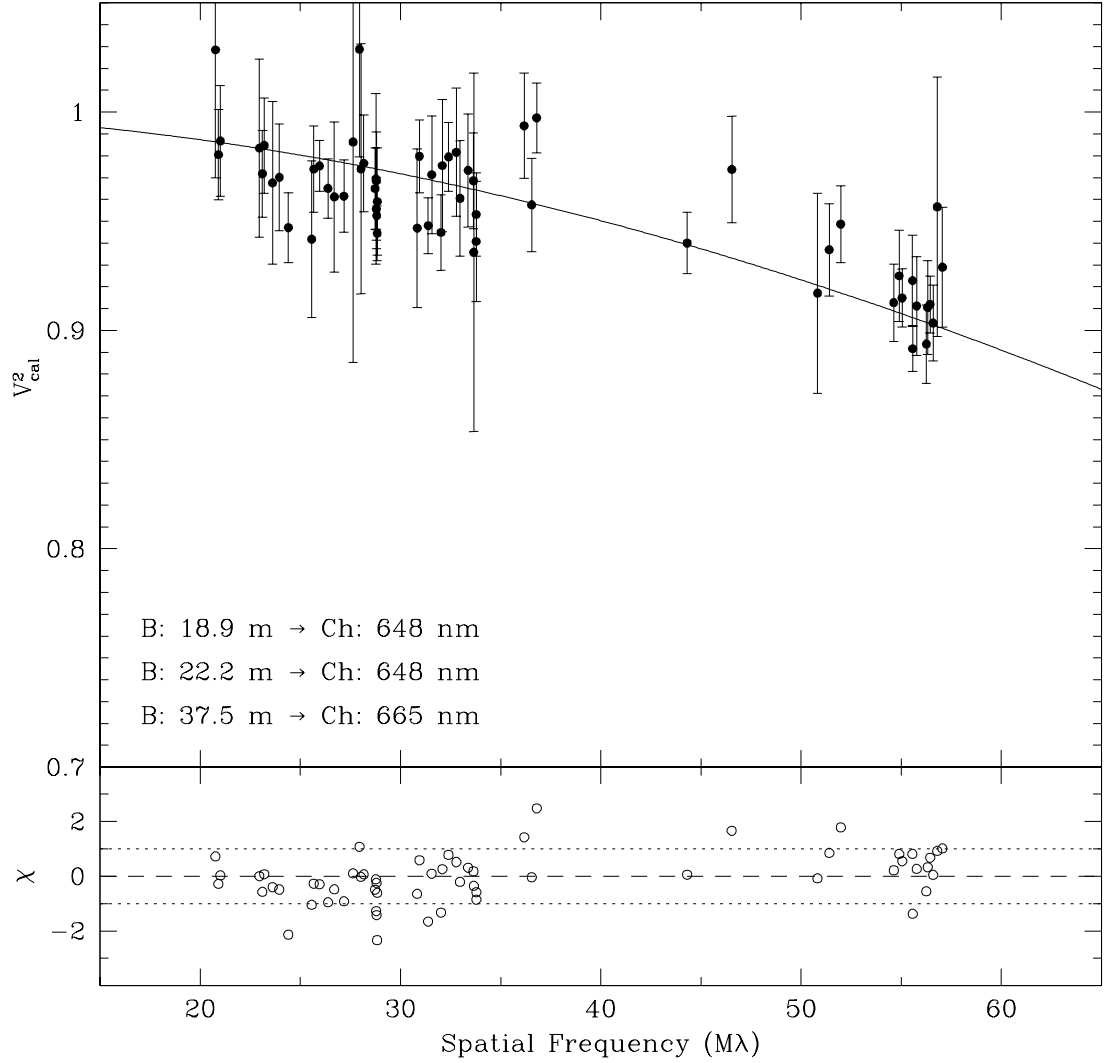}
\caption{Same as Fig.~\ref{fig:model_fit-BCMi}, but with the choice of
the spectral channels reversed.  The residuals in the lower panel are
calculated with respect to the uniform disk model representing the
stellar photosphere.  }
\label{fig:noHalpha-BCMi}
\end{figure}

To be consistent with the modeling of the H$\alpha$-emitting envelopes
of $\gamma$~Cas and $\zeta$~Tau~\citep{Tycner03,Tycner04}, we model
the envelopes of $\eta$~Tau and $\beta$~CMi using Gaussian models.  We
first fit a circularly symmetric Gaussian to the data by keeping the
axial ratio, $r$, fixed at unity and the resulting two-parameter fits
are listed in Table~\ref{tab:new_parameters}.  As expected, because
both stars have a relatively weak H$\alpha$ emission, the contribution
from the stellar photosphere to the net signal in the H$\alpha$
channel is large in both cases~(roughly 90\%).  The best-fit
circularly symmetric Gaussian model for $\eta$~Tau is shown in
Figure~\ref{fig:model_fit-ETau} along with the resulting normalized
residuals.  The majority of the data points do not deviate from the
model by more than one standard deviation, although the residuals in
Figure~\ref{fig:model_fit-ETau}~(which have a $\chi^2_{\nu}$ of 1.03)
show more variation than the residuals calculated from the difference
between the UD model and the data in the continuum channels~(see the
residuals in Figs.~\ref{fig:noHalpha-ETau}, \ref{fig:604nm_res-ETau},
and~\ref{fig:702nm_res-ETau} which have $\chi^2_{\nu}$ values of 0.83,
0.57, and 0.43, respectively).  Likewise the best-fit circularly
symmetric Gaussian model for $\beta$~CMi is shown in
Figure~\ref{fig:model_fit-BCMi}.

Fitting elliptical Gaussian models to the squared visibilities from
the H$\alpha$ channels of $\eta$~Tau and $\beta$~CMi yields model
parameters with an apparent ellipticity in both cases.  The model
parameters are listed in Table~\ref{tab:new_parameters} along with the
corresponding $\chi^2_{\nu}$ values, which also suggest that the
elliptical Gaussian models result in a slightly better fits.
Unfortunately, we cannot confidently rule out the possibility that the
apparent ellipticity is a result of systematic errors that could be
present in our data~(for example due to insufficient bias level
corrections).  However, it is interesting to note that
\citet{Quirrenbach97}, who observed $\eta$~Tau and $\beta$~CMi with
the Mark~III interferometer~(and thus were influenced by different
systematic errors), reported similar model parameters for both stars.

\section{The Data}
\label{sec:data}

The modeling presented in this paper and \citet{Tycner03,Tycner04}
has resulted in properties of H$\alpha$-emitting envelopes for a total
of four Be stars that have been observed with the NPOI.  Because we
model each H$\alpha$-emitting envelope in nearly the same way
as~\citet{Quirrenbach97}, we can also include their results for three
extra Be stars to increase the sample size.
Table~\ref{tab:be_star_sizes} lists the names and the spectral
types~\citep[from the Bright Star Catalogue, BSC,][]{Hoffleit82} of
the Be stars observed with the NPOI, along with those that have model
parameters based on the work of~\citet{Quirrenbach97}, at the bottom
of the list.  We see from the table that we have a range of spectral
types, and thus it should be possible to investigate not only the
relationship between the H$\alpha$ emission and the physical extent of
the emitting region, but also to test for any obvious dependencies on
spectral type.  We begin by discussing the sources of the
observational parameters derived from both interferometric and
spectroscopic observations of each star.

For the Be stars $\gamma$~Cas, $\eta$~Tau, $\zeta$~Tau, and
$\beta$~CMi, best-fit model parameters describing an elliptical
Gaussian model have been presented in \citet{Tycner03,Tycner04} and
\S~\ref{sec:models}.  The model parameters that are relevant to our
discussion are the angular size of the major axis~($\theta_{\rm mj}$)
and the ratio of the minor to major axis~($r$).  \citet{Quirrenbach97}
also obtained the same model parameters for $\phi$~Per, $\psi$~Per,
and 48~Per and therefore in Table~\ref{tab:be_star_sizes} we adopt
their parameters for these stars.

The radiation from the circumstellar disk of a Be star can be
characterized by the equivalent width~(EW) of the H$\alpha$ emission
line.  Unfortunately, our interferometric observations do not contain
this type of detailed spectroscopic information, and thus we need to
rely on independently obtained spectroscopic observations.  For Be
stars that are known to possess H$\alpha$ emission that is stable on
long time scales, it is not necessary to obtain spectra that cover the
same time-frame as the interferometric observations.  On the other
hand, for stars whose variability is well documented, spectroscopic
observations taken as close in time as possible to the interferometric
observations are required.  We have searched the literature for
spectroscopic information on the H$\alpha$ emission that would cover
the same period as our interferometric observations, which were
acquired between 1997 and 1999, and those of \citet{Quirrenbach97},
which were obtained throughout 1992.

To check for consistency and long-term stability of the H$\alpha$
emission of the targets observed with the NPOI, we have obtained
echelle spectra of $\gamma$~Cas and $\zeta$~Tau on 2003~December~9,
and $\eta$~Tau and $\beta$~CMi on 2004~March~8.  The echelle spectra
were obtained using the Lowell Observatory's Solar-Stellar
Spectrograph~(SSS) located at the same site as the NPOI.  The SSS
instrument and the IDL-based reduction package that was used to
extract the spectra from the raw echelle orders were described
by~\citet{Hall94}.  The resulting spectra of the H$\alpha$ region,
shown in Figure~\ref{fig:Halpha_4plots}, have a resolving power of
about 30,000 and SNR in the range of 150--200 at the continuum
wavelengths, and higher at the emission line.

\subsection{H$\alpha$ emission of $\eta$~Tau and $\beta$~CMi}

\begin{figure}
\plotone{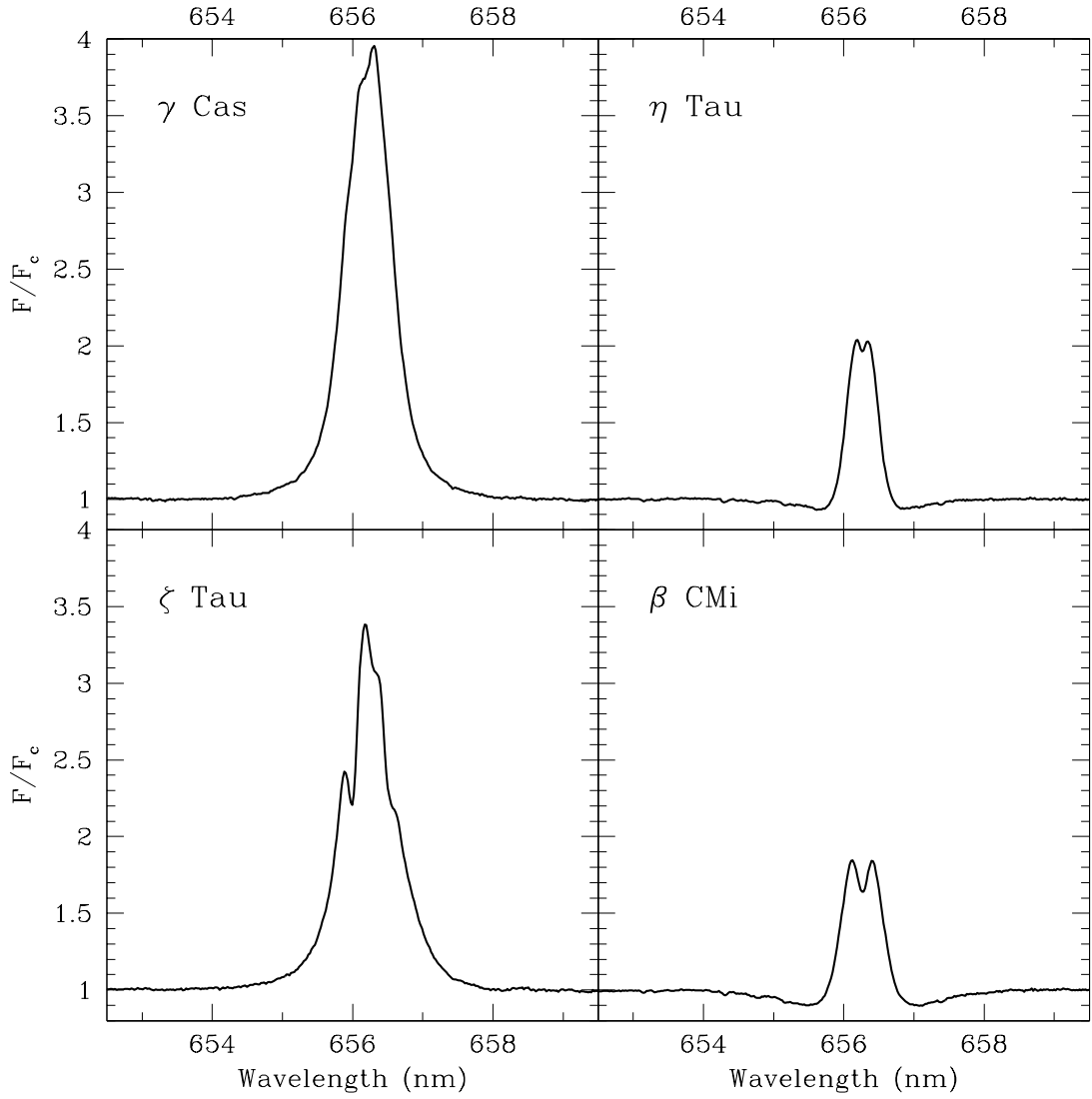}
\caption{The H$\alpha$ profiles of the four Be stars obtained with the
Solar-Stellar Spectrograph.  Each panel shows only a fraction of the
648--660~nm spectral region covered by the echelle order that
contained the H$\alpha$ line. }
\label{fig:Halpha_4plots}
\end{figure}

The H$\alpha$ emission profiles of $\eta$~Tau and $\beta$~CMi are
presented in the literature for a number of different epochs.  By
inspecting the profiles obtained by various observers between 1980 and
1999~\citep{Andrillat82,Slettebak92,Hanuschik96b,Banerjee00}, as well
as the profiles from 2004 shown in Figure~\ref{fig:Halpha_4plots}, we
conclude that the overall emission remained largely unchanged for both
stars over the period covered by the observations.  This is true not
only for the relative shape of the line but also the width and the
peak intensity.  The only exception are the observations reported by
\citet{Banerjee00}, where the emission appears higher above the
continuum (by $\sim$10 and $\sim$30~\% for $\beta$~CMi and $\eta$~Tau,
respectively) than in all other references.  Because
\citet{Banerjee00} have a spectral coverage of approximately 2.5~nm
centered on H$\alpha$, which is narrower than the full extent of the
underlying absorption line, their H$\alpha$ profiles are susceptible
to incorrect normalization of the continuum level when the absorption
line is not completely filled in (as is the case in $\eta$~Tau and
$\beta$~CMi).  Treating the extended wings of the absorption component
as the continuum level will then result in an overestimation of the
peak intensity of the emission component.  Therefore, we conclude that
the H$\alpha$ profiles of $\eta$~Tau and $\beta$~CMi do not change
significantly on long time scales and for these two stars we use the
equivalent widths of the emission lines observed in 2004, shown in
Figure~\ref{fig:Halpha_4plots}.

\subsection{H$\alpha$ emission of $\gamma$~Cas and $\zeta$~Tau}

Both $\gamma$~Cas and $\zeta$~Tau are known to show complex variations
in their H$\alpha$ emission profiles on time scales from months to
years, but in both cases the variations on the longer time scales are
always more significant.  Furthermore, because the EW measure is an
integrated quantity, it has the tendency to be insensitive to the
small-scale variations in the line profile, typically seen at the
shortest time scales.  Therefore we are only interested in the changes
in the EW of H$\alpha$ on the time scales of years.

In the case of $\gamma$~Cas the spectroscopic observations conducted
by~\citet{Banerjee00} are the closest in time to our interferometric
observations that were obtained between 1997 and
1998~\citep{Tycner03}. \citet{Banerjee00} reports H$\alpha$ EW of
$-2.25$~nm at the end of 1998, and $-2.67$~nm for observations
obtained a year later.  By comparing the above values with the EW of
$-2.57$~nm that we obtain for the H$\alpha$ profile of $\gamma$~Cas
from 2003~(recall Fig.~\ref{fig:Halpha_4plots}), we conclude that
variability in the EW at the $\sim$20\% level, on time scale of a
year, can be expected.  However, because the observations of
\citet{Banerjee00} from 1998 were acquired only one month after our
1998 interferometric observations, we use their reported EW of
H$\alpha$ as the corresponding emission measure for our observations.

The interferometric observations of $\zeta$~Tau that we have analyzed
in \citet{Tycner04} were obtained on 1999~March~1.  Once again, the
H$\alpha$ observations by \citet{Banerjee00} conducted on
1999~March~16 are the closest in time to our observations.  Because
the H$\alpha$ profile variations on time scales of days are very small
in $\zeta$~Tau~\citep{Hanuschik96b}, the EW of $-2.06$~nm reported by
\citet{Banerjee00} is suitable for our analysis.  It is also
interesting to note that this value is effectively the same as the EW
of $-2.09$~nm we obtain for the profile shown in
Figure~\ref{fig:Halpha_4plots}.  We should also note that in the case
of $\gamma$~Cas and $\zeta$~Tau the H$\alpha$ emission completely
fills in the underlying absorption component and therefore the
normalization of the continuum in these stars is less susceptible to
errors, like the ones discussed in the previous section.

\subsection{H$\alpha$ emission of $\phi$~Per, $\psi$~Per, and 48~Per}

By comparing the H$\alpha$ profiles of $\phi$~Per, $\psi$~Per, and
48~Per from 1980~\citep{Andrillat82}, 1989~\citep{Slettebak92},
1993~\citep{Hummel_W95}, and 1998--2000~\citep{Banerjee00}, we
conclude that only $\phi$~Per has significantly varying H$\alpha$
emission, but even in that case the main variation is a gradual
weakening of the emission over many years.  Since the Mark~III
observations of the above stars date back to 1992, we use the EW of
H$\alpha$ reported by~\citet{Hummel_W95}, who observed these stars in
1993.  The three stars, $\phi$~Per, $\psi$~Per, and 48~Per, have shown
equivalent widths of $-3.50$, $-3.37$, and $-2.23$~nm, respectively.

\section{The Analysis}

\subsection{Physical Extent of a Circumstellar Region}

Because distance has a direct effect on angular size, the angular
measurements of the major axes of the circumstellar regions must be
either transformed into physical sizes~(if the distances are known) or
normalized with respect to the angular diameters of the central
stars~(which must be estimated since they have not been measured
directly).  The latter method was used by \citet{Quirrenbach97}, and,
although it does eliminate the dependence on distance, it introduces a
dependence on the size of the central star.  Because we have a range
of spectral types, this approach is not optimal.

A much better approach is to use the known distances to obtain
directly the physical extent of each region.  We use the observed
Hipparcos parallax~($\pi$) to derive the corresponding distance~($d$)
to each star.  This in turn allows us to convert the best-fit value
for the angular size of the major axis of the elliptical Gaussian
model into a corresponding size in meters.
Table~\ref{tab:be_star_sizes} lists the angular size of the major
axis~($\theta_{\rm mj}$) of each H$\alpha$-emitting region and the
corresponding physical size in meters~($D_{\rm mj}$), along with the
Hipparcos data used to calculate the distance to each star.

\subsection{The Net H$\alpha$ Emission}
\label{sec:net_Halpha}

The equivalent widths of the H$\alpha$ emission lines typically
reported in the literature only measure the amount of emission above
the continuum level.  Because B-type stars possess significant
H$\alpha$ absorption profiles, we need to correct for the
``filling-in'' effect of the absorption line to recover the total
amount of emission.  In our approach we assume that the entire
absorption line is filled in by the superimposed emission.  This is a
good approximation for the sources considered here, where the
H$\alpha$ emission lines are stronger and wider than the underlying
absorption lines.  Even in the case of $\eta$~Tau and $\beta$~CMi,
where the absorption lines are not completely filled in~(recall
Fig.~\ref{fig:Halpha_4plots}), the above assumption can still be
applied, because our EW measures of the emission are already reduced
by the amount the absorption line is not filled in.

To obtain an estimate on the EW of the absorption line, we follow the
procedure developed by~\citet{Cote87}, who modeled the relationship
between the EW of the H$\alpha$ absorption line and the intrinsic
$(B-V)_0$ color index for normal main sequence B-type stars.  This can
be done because in B-type stars the strength of the H$\alpha$
absorption line increases monotonically with spectral type to a
maximum at type A0.  We should also note that for stars in our sample
that are not on the main sequence, the equivalent widths of the
absorption lines derived using the above method will be slightly
overestimated~(by not more than 0.1~nm).  Because this is at the same
level as the intrinsic scatter observed in the H$\alpha$ equivalent
widths for stars of the same spectral type~\citep[see, for example,
Fig.~1 in][]{Cote87}, we neglect this overestimation in our analysis.

In Table~\ref{tab:be_star_mag} we list the observed $B-V$ colors from
the BSC for each of the stars.  Deriving the intrinsic $(B-V)_0$ color
of the central star from the observed $B-V$ value is not trivial, even
if the corrections for the interstellar extinction are known, because
Be stars can still appear intrinsically redder than normal B-type
stars of the same spectral type.  This is typically attributed to
free-bound and free-free emission, as well as a possible circumstellar
reddening~\citep{Schild78}.  For this reason we estimate the $(B-V)_0$
index based on the spectral type and not on the observed $B-V$ color.
Table~\ref{tab:be_star_mag} lists the $(B-V)_0$ values based on the
tabulation of~\citet{FitzGerald70} which in turn is used to calculate
the equivalent widths of the H$\alpha$ absorption lines~\citep[via
eq.~6 of][]{Cote87}.  These values are then combined with the
equivalent widths of the observed H$\alpha$ emission lines~($W_{{\rm
H}\alpha}$) to give a measure of the net H$\alpha$ emission~($E_{{\rm
H}\alpha}$) from each circumstellar region~(both quantities are listed
in Table~\ref{tab:be_star_emission}).

\subsection{The H$\alpha$ Luminosity}

The net H$\alpha$ emission derived in previous section, although
corrected for the ``filling-in'' effect of the absorption line, is
still measured with respect to the underlying stellar continuum.
Because we are dealing with spectral classes ranging from B0 to B8, we
need to account for the different continuum flux levels.  We use two
independent methods to estimate the stellar continuum level.  One
method is based on the photometry and thus requires us to account for
the different distances to the sources.  The other method, which is
distance independent, uses model energy distributions to represent the
underlying stellar continuum.  Because the second method does not
account for the continuum emission from the circumstellar region, as
well as requires an estimate of the stellar radius, it is only used to
verify the photometrically derived results.

\subsubsection{H$\alpha$ luminosity based on photometrically determined continuum}

Obtaining the flux levels of the individual stars from photometry is
complicated by the interstellar extinction.  Using the observed $B-V$
color and the estimated intrinsic $(B-V)_0$ color based on the
spectral type, we can derive the color excess $E(B-V)$.  This approach
should work for the later type Be stars, but will tend to overestimate
the color excess~(by possibly as much as 0.15~mag) for spectral types
B3e and earlier, because of the intrinsic reddening due to the
free-bound and free-free emission from the circumstellar
envelope~\citep{Schild78,Schild83}.  Fortunately, there are other
methods that can be used to estimate the color excess due to the
interstellar extinction.  For example, \citet{Beeckmans80} used the
depth of the interstellar absorption bump at 2200~\AA\ to derive the
values of $E(B-V)$ for a number of Be stars and in
Table~\ref{tab:be_star_mag} we list their values of $E(B-V)$ for all
but two of our stars.  The two missing Be stars are $\eta$~Tau and
$\beta$~CMi, and because these stars have the latest spectral type, as
well as being the closest, we simply derive an estimate of $E(B-V)$
based on their observed $B-V$ colors and the intrinsic $(B-V)_0$
colors derived from the spectral type.

With a known color excess $E(B-V)$ we can correct the apparent $V$
magnitude of each star for interstellar extinction using:
\begin{equation}
V_0 = V - A_V  = V - R_V \cdot E(B-V),
\end{equation}
where $R_V$ is the reddening constant, defined as $A_V / E(B-V)$. The
typical value of $R_V$ found by~\citet{Schultz75} is 3.14$\pm$0.10.
Similarly, the observed $V-R$ colors~(see Table~\ref{tab:be_star_mag})
can be corrected for interstellar reddening using the ratio of
$E(V-R)/E(B-V) = 0.78\pm0.02$ derived by~\citet{Schultz75}, and in
turn the $R_0$ magnitudes can be obtained.  The $R$-band has an
effective central wavelength of $\sim$658~nm and a FWHM of
$\sim$138~nm, and thus $R_0$ magnitude can be used to measure the
continuum flux level in the H$\alpha$ region.  Also, the relatively
large spectral width of the $R$-band allows us to neglect the
contribution of the H$\alpha$ emission to $R_0$.

To convert a $R_0$ magnitude into a corresponding flux above Earth's
atmosphere~($f_{\star}$) we use
\begin{equation}
f_{\star} = f_{\rm s}  10^{-0.4 (R_{\star}-R_{\rm s})},
\end{equation}
where $f_{\rm s}$ and $R_{\rm s}$ are the mean flux density~(in the
$R$-band) and the $R$ magnitude of a standard star, respectively.
Because we are interested in the functional dependence of the net
H$\alpha$ emission on the disk size, the actual value of $f_{\rm s}$
is not important.  Nevertheless, in order to work with values that
have physically meaningful units we use a \emph{representative} value
for the absolute flux density of $f_{\rm s} =
1.76\times10^{-11}$~W~m$^{-2}$nm$^{-1}$ defined for a standard star
with $R_{\rm s}=0$~\citep{Johnson66a}.  The values for $f_{\star}$
based on the $R_0$ magnitudes are listed in
Table~\ref{tab:be_star_lum}, along with the corresponding \emph{total}
radiation at H$\alpha$~($4\pi d^2 f_{\star}$).

Finally to work with an H$\alpha$ emission measure that is independent
of distance we multiply the net H$\alpha$ emission, $E_{{\rm
H}\alpha}$, derived in previous section, by the total radiation in the
H$\alpha$ region, to obtain
\begin{equation}
\label{eqn:Halpha_lum}
L_{{\rm H}\alpha} = 4\pi d^2 f_{\star} E_{{\rm H}\alpha},
\end{equation}
which we refer to as the H$\alpha$ luminosity.  The values for the
H$\alpha$ luminosities are given in Table~\ref{tab:be_star_emission}
along with their corresponding uncertainties, which account for the
errors in the distance determination~(largest source of uncertainty),
photometry, and the H$\alpha$ equivalent widths, but do not include
the uncertainty associated with the determination of the absolute flux
density of a standard star.

\subsubsection{H$\alpha$ luminosity based on synthetic spectra}
\label{sec:synthetic}

\begin{figure}
\plotone{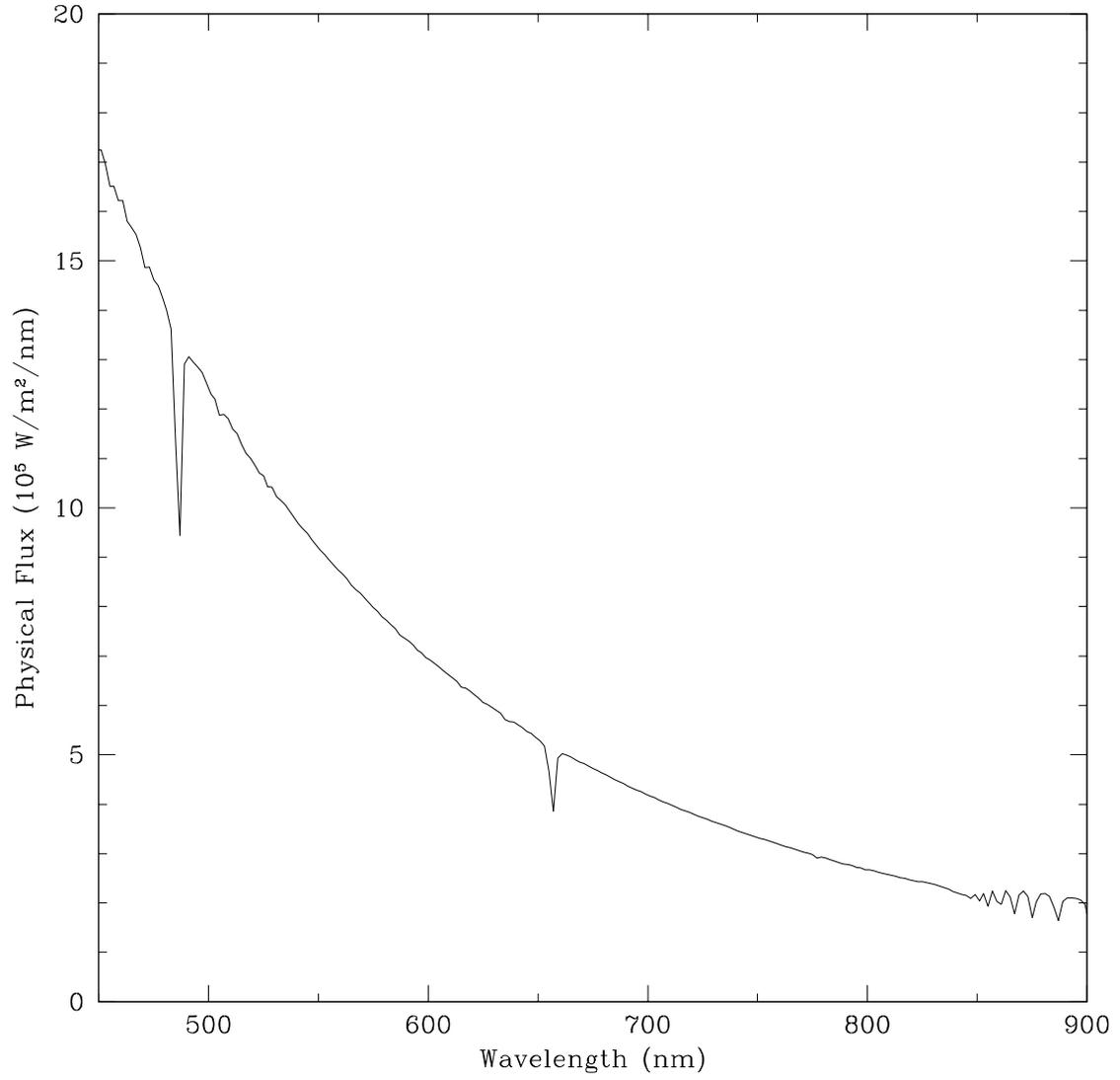}
\caption{Synthetic energy distribution for $T_{\rm eff}=12000$~K and
$\log g = 4.0$.  }
\label{fig:Spec_12000K}
\end{figure}

Instead of using photometry to derive the stellar continuum flux level
in the H$\alpha$ region, we can use synthetic spectra calculated for
atmospheric models of varying effective temperatures.  For this
purpose we have used the well known grid of stellar atmosphere model
spectra computed using the ATLAS9 program~\citep{Kurucz93}.  The
models were chosen to have effective temperatures that were the same
as, or as close to those estimated based on the spectral type~(see
Table~\ref{tab:be_star_lum}).  Because the grid covered the range of
the effective temperatures in approximately 500 and 1,000~K steps, for
some stars it was necessary to interpolate the fluxes in effective
temperature.  All models correspond to solar abundances, $\log g$ of
4~(for luminosity class V or IV) and 3.5~(for luminosity class of
III), although little variation between the two gravities is present,
and the standard microturbulence of 2~km~s$^{-1}$ was used.
Figure~\ref{fig:Spec_12000K} illustrates a sample of the physical
flux~($\mathcal{F}_{\star}$) as a function of wavelength obtained for
a 12,000~K model.

To estimate the total stellar radiation from the physical flux in the
H$\alpha$ region we need to account for the surface area of each star.
We estimate the radius of each star based on its spectral
type~\citep{Underhill79}, and the resulting synthetic total radiation
in the H$\alpha$ region~($4\pi R_{\star}^2 \mathcal{F}_{\star}^{\rm
S}$) is given in Table~\ref{tab:be_star_lum}.  It is evident from the
table that the synthetic values are systematically smaller than the
photometrically derived ones, and this should be expected because the
synthetic spectra do not account for the bound-free and free-free
emission from the circumstellar material.  Although there does exist a
general correlation between the net H$\alpha$ emission and IR excess,
a large intrinsic scatter is also observed between different sources,
and therefore it is not trivial to account for this extra
emission~\citep{Cote87,vanKerkwijk95}.  Furthermore, the continuum
level can be affected by the stellar rotation, especially when the
star is rotating close to its critical velocity~\citep{Townsend04}.
The assigned spectral type might not account fully for such deviation
from the continuum level that would be observed in the case of a
non-rapidly rotating star of the same spectral type.  Nevertheless,
because we are only interested in the synthetic continuum level as a
check for the photometrically derived values, we neglect the
contribution to the continuum level from the circumstellar region and
any possible changes due to the stellar rotation.  Taking the same
approach as in equation~(\ref{eqn:Halpha_lum}) we can calculate a
synthetic H$\alpha$ luminosity using
\begin{equation}
L_{{\rm H}\alpha}^{\rm S} = 4\pi R_{\star}^2 \mathcal{F}_{\star}^{\rm
S} E_{{\rm H}\alpha},
\end{equation}
where $E_{{\rm H}\alpha}$ is the net H$\alpha$ emission discussed in
\S~\ref{sec:net_Halpha}.  The $L_{{\rm H}\alpha}^{\rm S}$ values for
all seven stars are listed in Table~\ref{tab:be_star_emission}.

\section{Discussion}

\subsection{The H$\alpha$ Emission as a Tracer of the Circumstellar Disk}
\label{sec:Halpha-tracer}

\begin{figure}
\plotone{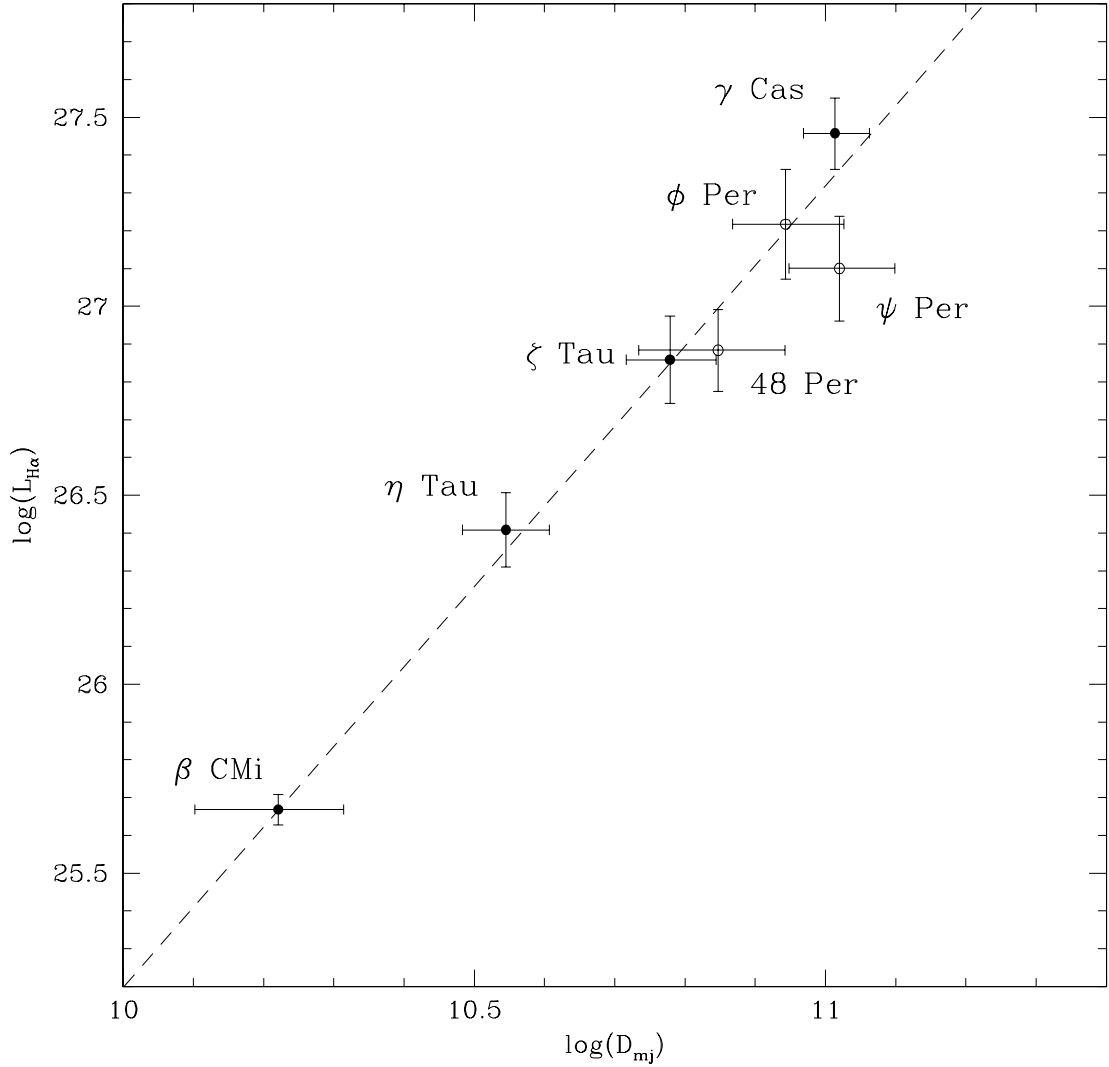}
\caption{The H$\alpha$ luminosity as a function of the size of the
major axis of the disk.  The circumstellar envelopes that have been
resolved by the Mark~III interferometer~\citep{Quirrenbach97} are
marked with {\it open circles}.  A linear fit in the log-log plot
produces a best-fit slope of 2.12$\pm$0.24~({\it dashed line}).  }
\label{fig:Halpha_Lum}
\end{figure}

We begin by looking at the relationship between the H$\alpha$ emission
and the physical extent of the circumstellar region.  Using the values
for the H$\alpha$ luminosity from Table~\ref{tab:be_star_emission} and
the physical size of the major axis from
Table~\ref{tab:be_star_sizes}, we construct a corresponding log-log
plot of $L_{{\rm H}\alpha}$ versus $D_{\rm mj}$ in
Figure~\ref{fig:Halpha_Lum}.  We obtain a Pearson correlation
coefficient of 0.98 and thus a definite correlation between the
parameters exists.  Fitting a line to the data in the log-log plot
yields a slope of 2.12$\pm$0.24 and a $\chi^2_{\nu}$ of 0.75.

The above result suggests that the emission is proportional to the
area of the H$\alpha$-emitting region independently of the different
inclination angles at which these systems are viewed~(although the
range of possible inclination angles represented by our small sample
of stars is not that large).  If the circumstellar region is optically
thin to H$\alpha$ emission then this might simply mean that the
\emph{emission measure}\footnote{Because the H$\alpha$ emission is
created by recombinations, the total emission in the optically thin
case is proportional to the volume integrated squared electron number
density, i.e. $\int N_{\rm e}^2 \dif V$, which is also known as the
emission measure~\citep[see, for example,][]{Millar00}.} of the disk,
which is proportional to the emitting volume, scales as the surface
area of the emitting region.  However, the study of energetics of Be
star envelopes by~\citet{Millar00} shows that the H$\alpha$ emission
originating in the equatorial plane of the circumstellar disk,
especially close to the central star~(for $r < 15R_{\star}$), is
optically thick.  Because this is the same region that is responsible
for the majority of the line emission, and if the disk has a constant
surface brightness, then the H$\alpha$ luminosity might be expected to
be proportional to the effective surface area of the circumstellar
disk.

\begin{figure}
\plotone{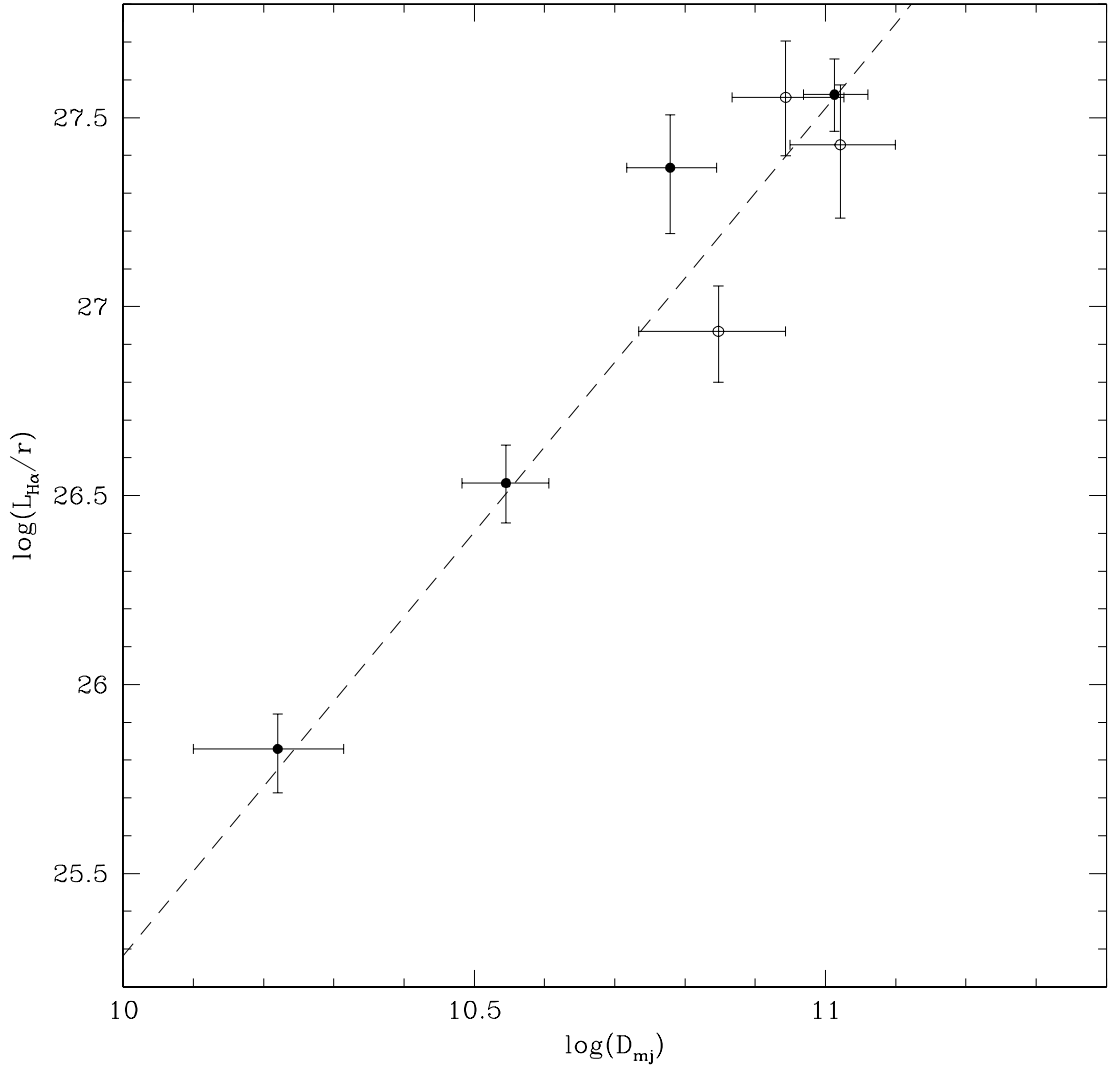}
\caption{Same as Fig.~\ref{fig:Halpha_Lum}, except that the H$\alpha$
luminosity has been divided by the axial ratio $r$.  A linear fit in
the log-log plot produces a best-fit slope of 2.24$\pm$0.27~({\it
dashed line}). }
\label{fig:all_Be_stars}
\end{figure}

If we assume that the circumstellar region is a geometrically thin
axisymmetric disk, then the axial ratio, $r$, we have obtained from
the elliptical Gaussian model can be used to estimate the inclination
angle $i$ between the direction perpendicular to the plane of the disk
and the line of sight~(i.e., $r \sim \cos i$).  Because the effective
area of a disk viewed at an angle $i$ can be expressed as $A\cos i$
where $A$ is the surface area, and if the H$\alpha$ luminosity is
proportional to the effective area, dividing $L_{{\rm H}\alpha}$ by
$r$ should remove any dependence on the inclination angle.
Figure~\ref{fig:all_Be_stars} shows the relationship between $L_{{\rm
H}\alpha}/r$ and $D_{\rm mj}$ where again the correlation coefficient
is high~(0.96) and the best-fit line has a slope of 2.24$\pm$0.26 and
$\chi^2_{\nu}$ of 1.19.  Based on this result we conclude that the
H$\alpha$ luminosity (corrected for the projection effect) is
consistent with an optically thick emission from a geometrically thin
disk of a relatively constant surface brightness.

Optically thick disks with no temperature gradients will have a
constant surface brightness as a function of radius.  Although actual
disks will have a range of temperatures, assigning a constant
temperature to the H$\alpha$-emitting region might be appropriate.
This is because~\citet{Millar99c}, who, by considering the rates of
energy gain and loss in the disks of four Be stars~(including
$\gamma$~Cas, $\psi$~Per, and $\beta$~CMi), have shown that the disk
temperature in the equatorial plane is fairly constant as a function
of radial distance, especially in the regions close to the central
star~(i.e., where the majority of the H$\alpha$ flux is expected to
originate).  Furthermore, our results suggest that the surface
brightness in the H$\alpha$-emitting region is not only constant as a
function of radial distance, but that it does not vary significantly
for stars of different effective temperatures.  This is somewhat
surprising, considering that the disk models of \citet{Millar99c} for
stars of different effective temperatures do not have the same
temperature profiles as function of radial distance, although in the
inner part of the disks~($<10 R_{\star}$) the temperature differences
are not that large and are in the 3,000--5,000~K range~(depending on
the distance from the equatorial plane).  It is also possible that
because the H$\alpha$ emission is optically thick in these inner
regions~\citep{Millar00}, the similar surface brightnesses for disks
with slightly different temperatures is a result of similar conditions
in the envelopes where the optical depth is roughly unity~(i.e., where
most of the radiation originates).

\begin{figure}
\plotone{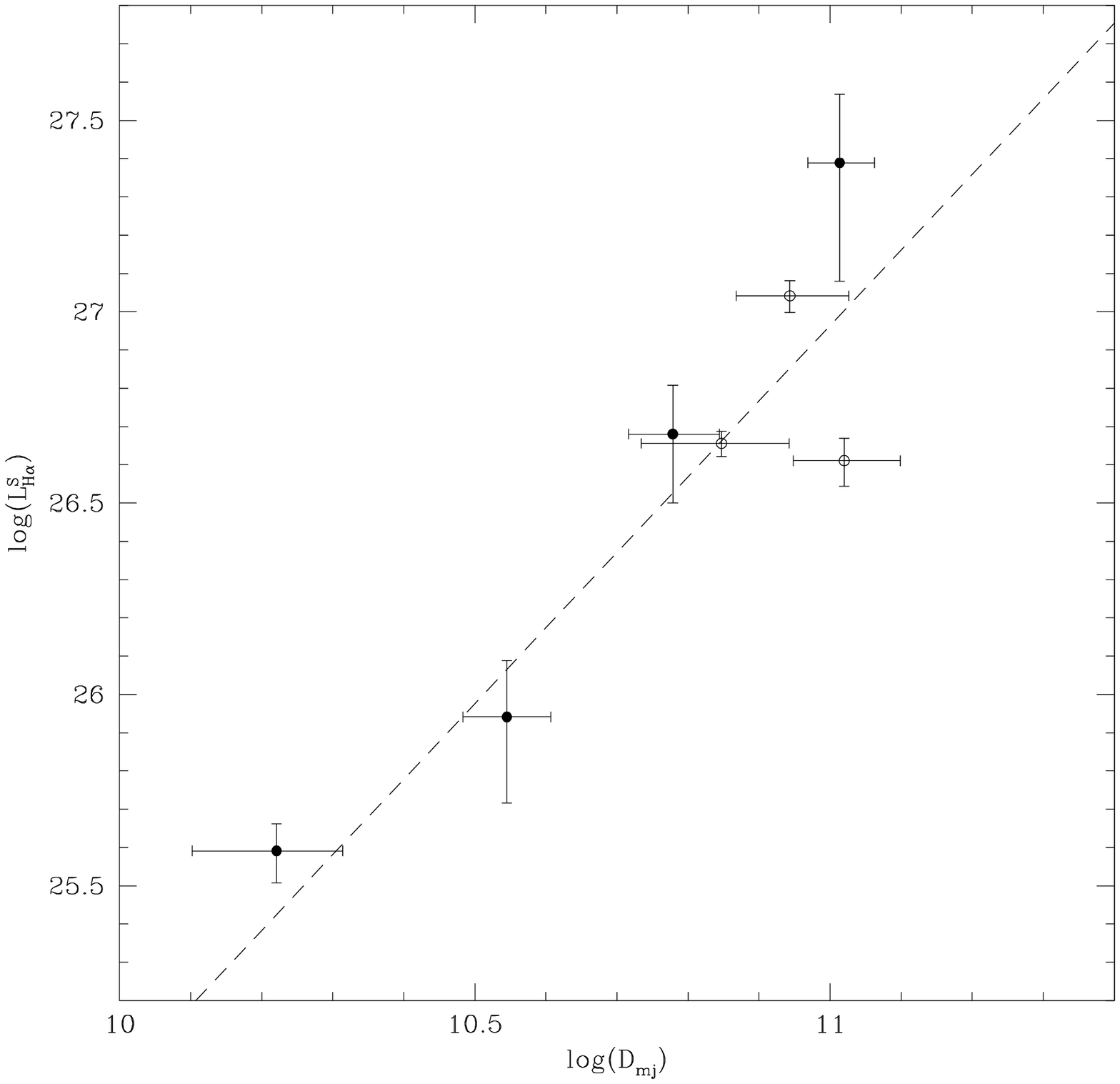}
\caption{Same as Fig.~\ref{fig:Halpha_Lum}, except that the H$\alpha$
luminosity has been calculated with respect to a synthetic continuum
level.  A linear fit in the log-log plot produces a best-fit slope of
1.98$\pm$0.33~({\it dashed line}).  }
\label{fig:NEW_HvsD}
\end{figure}

\begin{figure}
\plotone{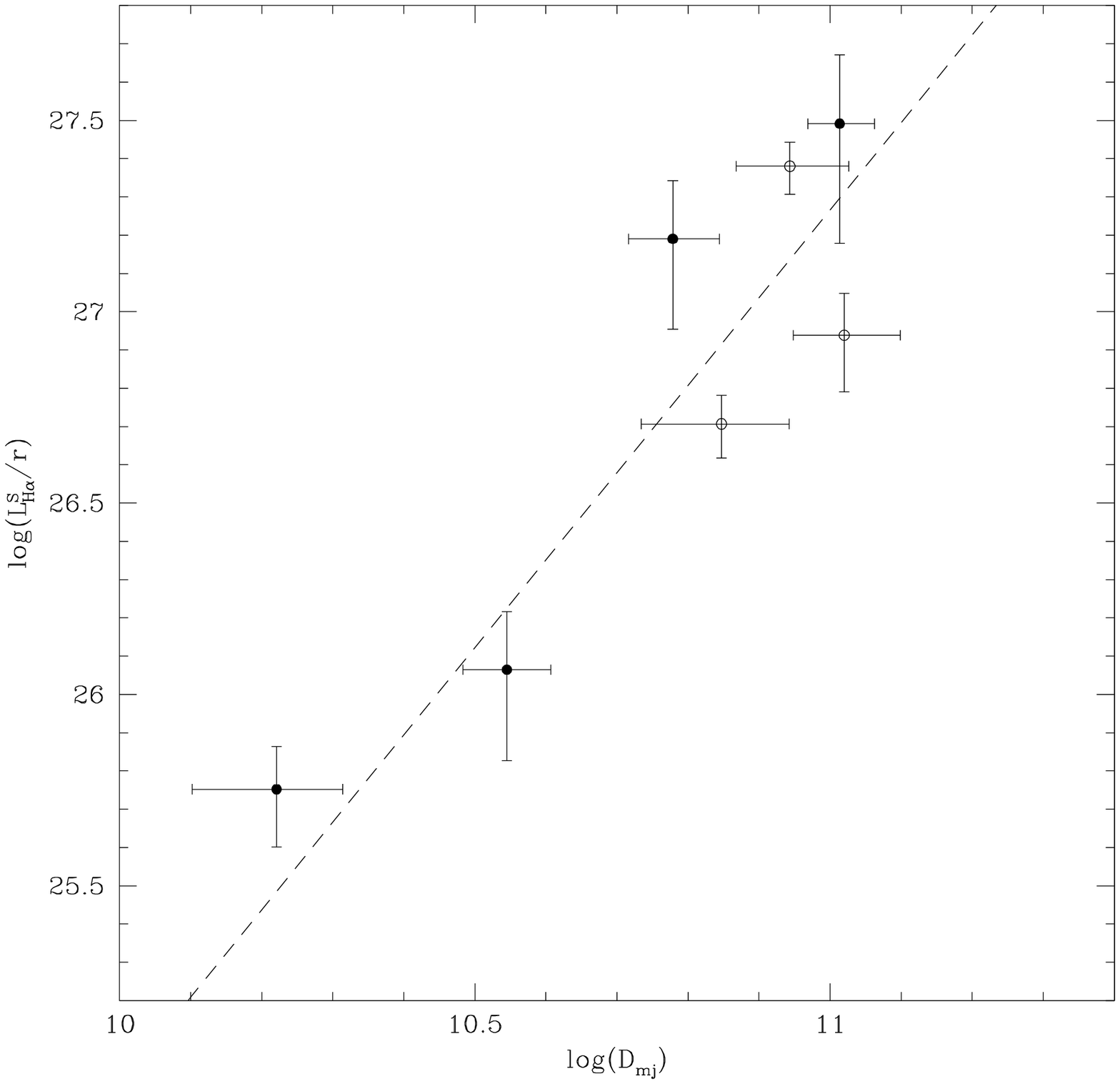}
\caption{Same as Fig.~\ref{fig:all_Be_stars}, except that the
H$\alpha$ luminosity has been calculated with respect to a synthetic
continuum level.  A linear fit in the log-log plot produces a best-fit
slope of 2.28$\pm$0.38~({\it dashed line}).  }
\label{fig:NEW_HvsD_w_r}
\end{figure}

Lastly, we need to verify that the visible trend in
Figure~\ref{fig:Halpha_Lum}~(as well as Fig.~\ref{fig:all_Be_stars})
is not a result of the functional dependence of $D_{\rm mj}$ and
$L_{{\rm H}\alpha}$ on the same determination of distance $d$, which
by itself will tend to introduce correlation between the two
parameters.  To verify that the correlations are not caused by the
dependence on $d$ alone, we use the synthetic H$\alpha$
luminosities~(recall \S~\ref{sec:synthetic}), which are measured with
respect to the synthetic continuum levels and thus are independent of
$d$, and we analyze the relationships between $L_{{\rm H}\alpha}^{\rm
S}$ and $L_{{\rm H}\alpha}^{\rm S}/r$ versus $D_{\rm mj}$ in
Figures~\ref{fig:NEW_HvsD} and \ref{fig:NEW_HvsD_w_r}, respectively.
Because only the abscissa depends on the distance in this case, a
trend can no longer be caused by the functional dependence on
distance.  These relations yield best-fit slopes of 1.98$\pm$0.33 and
2.28$\pm$0.38~(and correlation coefficients of 0.91 and 0.96), in
complete agreement with our previously determined values.  The
$\chi^2_{\nu}$ values for the relations based on the synthetic
continuum levels are slightly larger~(approximately twice as large)
than before, but this might be solely related to the unaccounted
systematic uncertainties associated with the estimated radii, as well
as with the fact that the synthetic spectra do not account for the
continuum emission from the circumstellar regions.

\subsection{Disk Size and Stellar Properties}

\subsubsection{Size versus Spectral Type}

\begin{figure}
\plotone{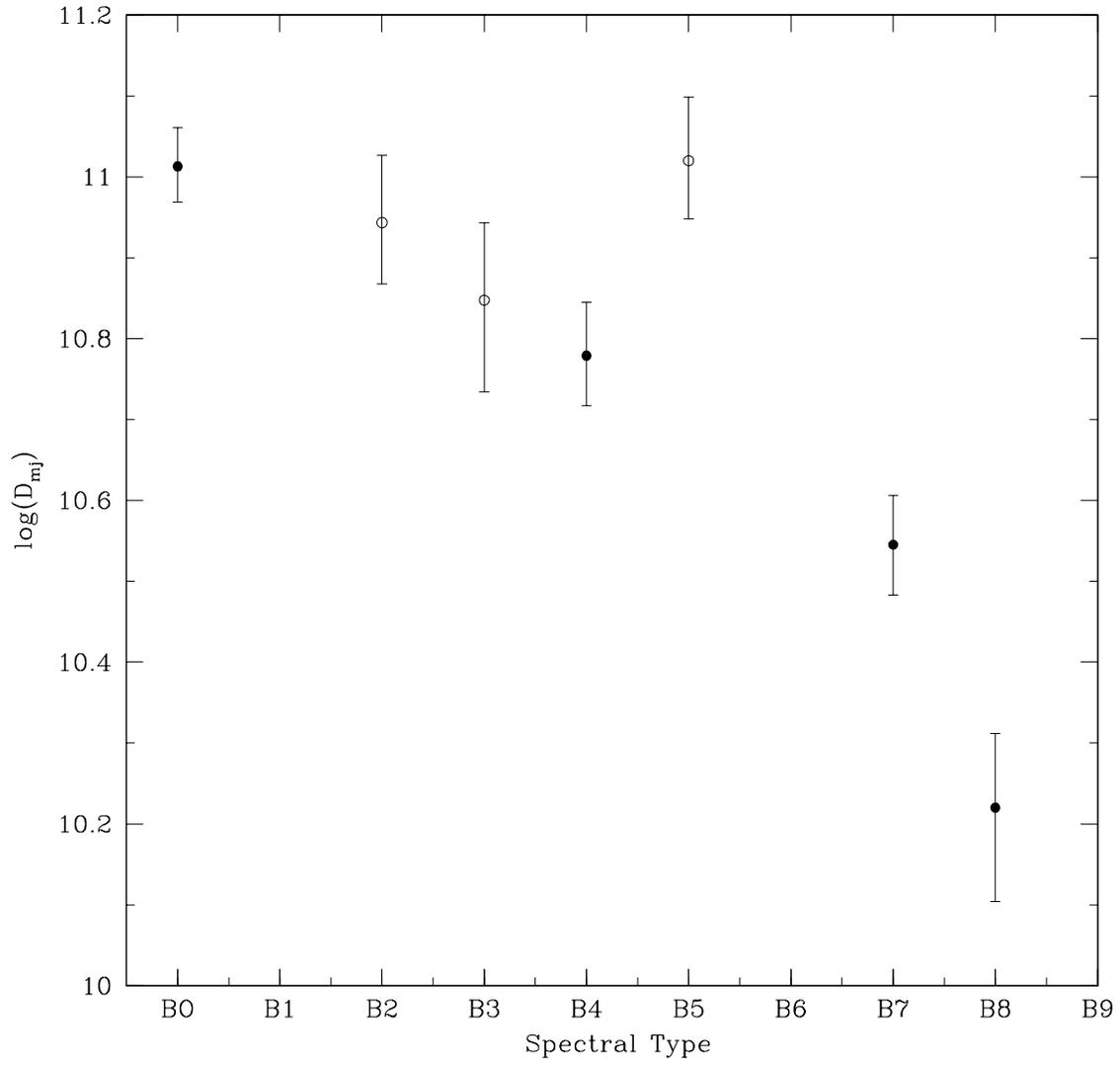} 
\caption{The size of the major axis of the disk as a function of the
spectral type.}
\label{fig:size_vs_ST}
\end{figure}

\begin{figure}
\plotone{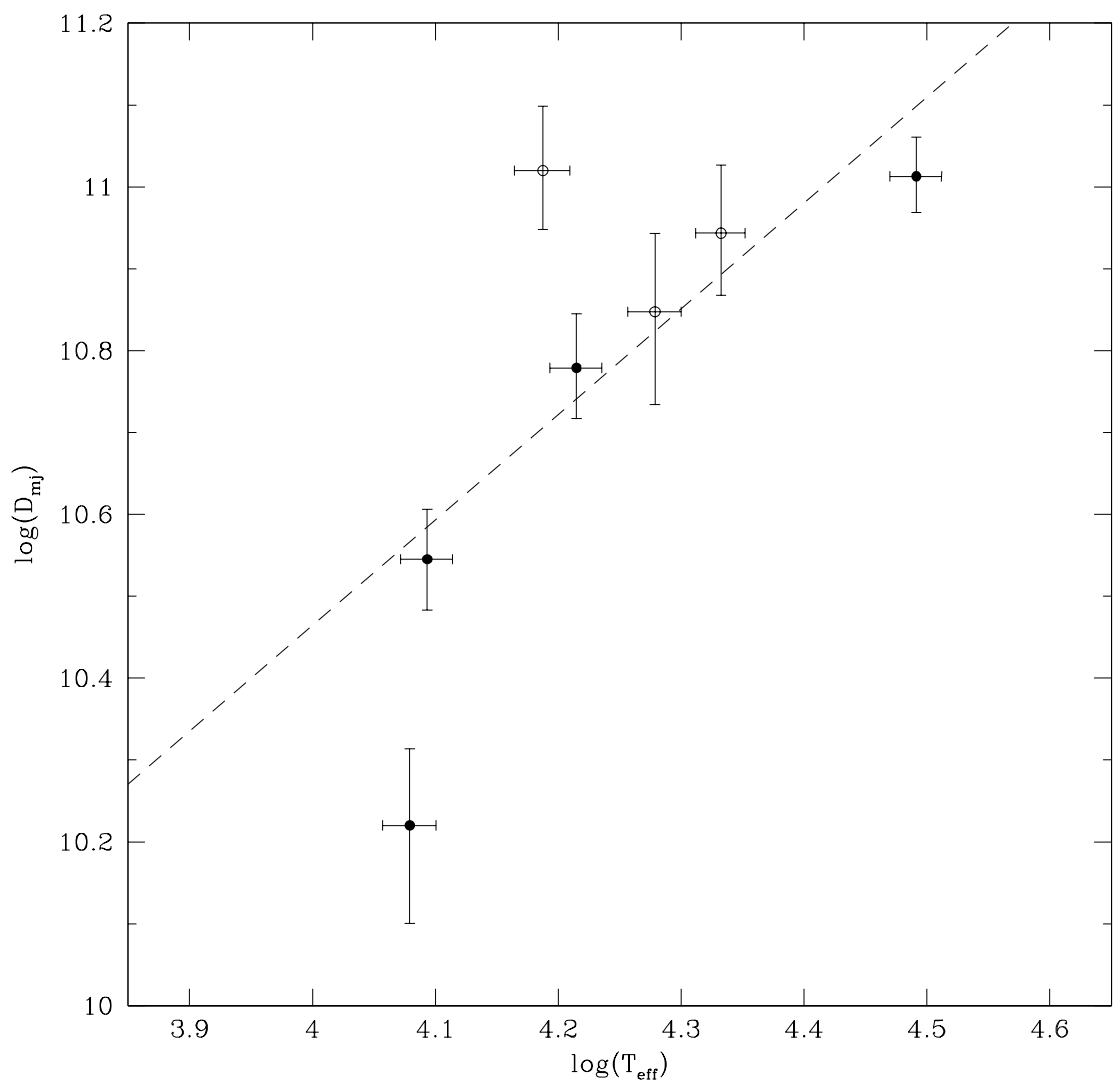}
\caption{The size of the major axis of the disk as a function of the
effective temperature of the central star. }
\label{fig:size_vs_temp}
\end{figure}

The origin of the circumstellar regions around Be stars is still
debated.  One of the simpler tests is to search for a possible
correlations between the size of the circumstellar region and the
spectral type of the central star.  Figure~\ref{fig:size_vs_ST}
illustrates the relationship between the physical extent of the major
axis and the spectral type for the seven stars listed in
Table~\ref{tab:be_star_sizes}.  Within our small sample of stars, it
appears that the size of the H$\alpha$-emitting region decreases
toward later spectral type.  This is also shown in
Figure~\ref{fig:size_vs_temp}, which plots $D_{\rm mj}$ versus the
effective temperature from Table~\ref{tab:be_star_lum}.  The visible
correlation is not nearly as strong as the one found for the H$\alpha$
emission and the size of the emitting region, but it still yields a
correlation coefficient of 0.75 and a slope of 1.29$\pm$0.19.
Unfortunately, the small number statistics make these results
tentative, but with future observations of a larger set of stars, it
will be possible to explore this relationship further.

If the apparent dependence of $D_{\rm mj}$ on the spectral type is
real, it is possible that this is due to the tendency of hotter stars
to have physically larger circumstellar regions, but equivalently this
can also be attributed to a simple ionization effect.  For example,
\citet{Millar99a} investigated models of circumstellar regions of Be
stars with a range of effective temperatures and concluded that in the
case of the hot Be star $\gamma$~Cas, the circumstellar envelope is
almost completely ionized, in contrast to the cool Be stars~(like
$\beta$~CMi), whose envelopes in and near the equatorial plane~(where
most of the emission normally originates) is mostly neutral~\citep[see
also][]{Millar98,Millar99b}.  This suggests that the cooler Be stars
might have smaller H$\alpha$-emitting regions due to the
ionization-bounded conditions in their circumstellar regions.

\subsubsection{Size versus Rotational Velocity}

Another commonly examined relationship is the one between the size of
the region and the rotational velocity.  The connection between the
two would be expected if rotation has a direct effect on the disk
formation, as originally suggested by~\citet{Struve31}.  In
Table~\ref{tab:be_star_rotation} we list the $v\sin i$ values
from~\citet{Slettebak82}, the estimated\footnote{The rotational
velocity is estimated by assuming that the circumstellar envelope is a
thin axisymmetric disk that is aligned with the stellar equator, in
which case the axial ratio, $r$, can be used to estimate $i$.}
rotational velocities~($v_{\rm est}$), along with the critical
velocities from~\citet{Porter96}.  We find no correlation between the
physical extent of the H$\alpha$-emitting region and the observed
$v\sin i$ value.  This is in agreement with previous studies that
concluded that there is no correlation between the strength of the
H$\alpha$ emission and the $v\sin i$ value.  

\begin{figure}
\plotone{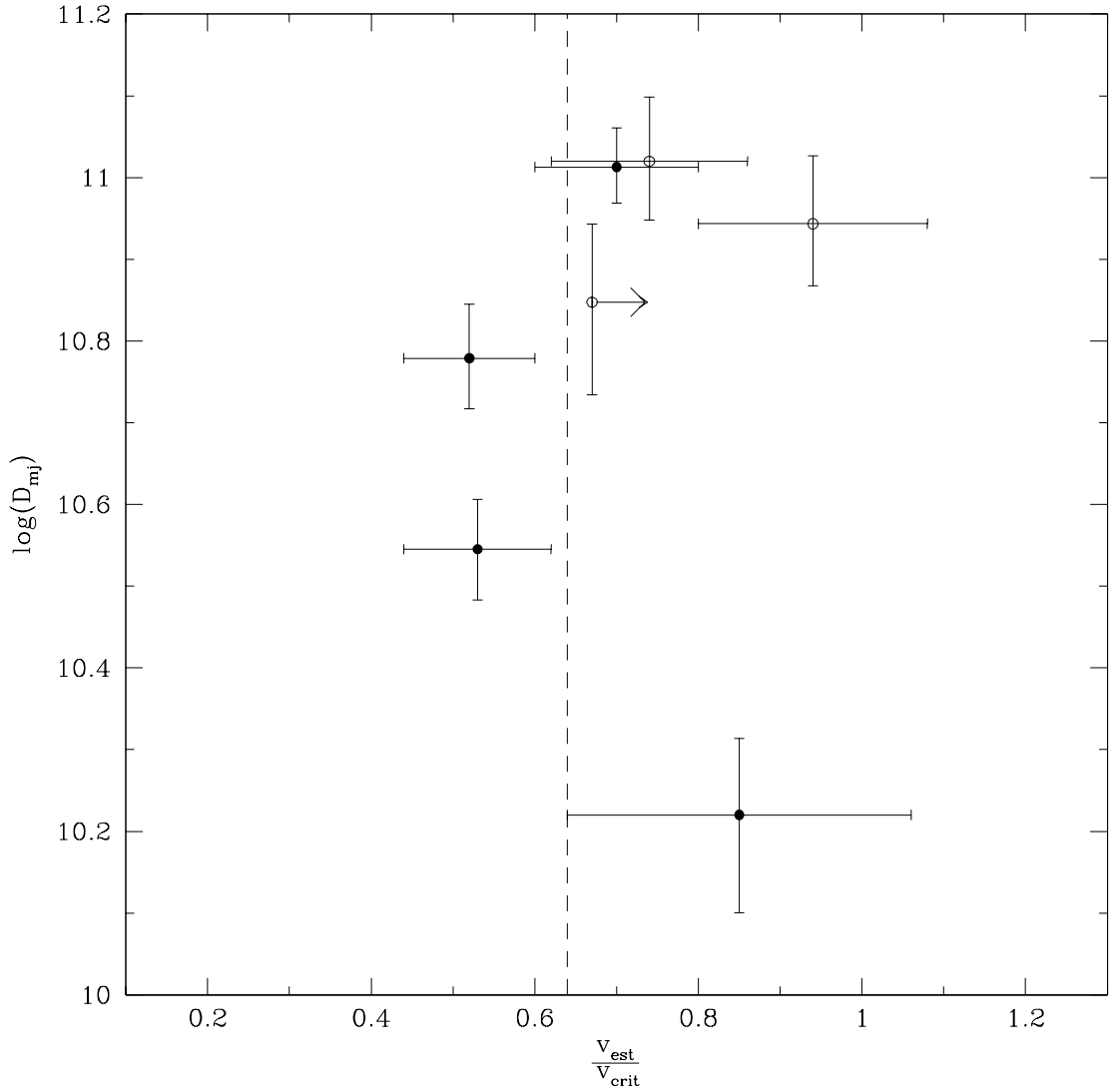}
\caption{The the size of the major axis of the disk as a function of
the estimated rotational velocity~(in units of critical velocity).
The weighted average value of $v_{\rm est}/v_{\rm crit}$ is shown with
{\it dashed line}.  }
\label{fig:size_vs_v}
\end{figure}

Because we derive an inclination angle $i$ from the apparent
ellipticity of the H$\alpha$-emitting disk, our estimates on the
rotational velocities of Be stars are quite independent of the
estimates that are obtained for systems that are thought to be viewed
edge-on~(i.e., using Be-shell stars).  Therefore our estimate on the
average rotational velocity, expressed as a fraction of $v_{\rm
crit}$, should be independent of similar estimates obtained from
observations of Be-shell stars.  In Table~\ref{tab:be_star_rotation}
we list the values of the ratio of $v_{\rm est}$ to $v_{\rm crit}$ for
all stars, including 48~Per for which only a lower limit can be
obtained~(see also Fig.~\ref{fig:size_vs_v}).  We obtain a weighted
average~(excluding 48~Per) of 0.64$\pm$0.14 for $v_{\rm est}/v_{\rm
crit}$, which is in excellent agreement with the work of
\citet{Porter96}, who concluded that the distribution of the
rotational velocities is sharply peaked at $\sim$0.7$v_{\rm crit}$.

Assuming that the $v\sin i$ values obtained from spectral lines are
not underestimated~(for example due to the equatorial gravity
darkening caused by rapid rotation), the above result suggests that Be
stars are rotating well below their critical limits.  On the other
hand, if the $v\sin i$ values are systematically underestimated by
tens of percent~\citep[as suggested by][]{Townsend04}, then it is
still possible that these stars have near-critical rotation rates.  A
much more direct method for determining the rotational velocity of
stars is the observation of the rotationally induced distortion of the
stellar photosphere by long-baseline interferometry.  In fact, the
first result suggesting a near-critical rotation in the case of one Be
star already exists~\citep{Domiciano03}.  It is anticipated that with
the advent of optical interferometric observations at baselines of
hundreds of meters~(such as those of the NPOI), further observational
constraints on the stellar rotation and its role in Be disk formation
will be possible.

\section{Conclusions}

Our analysis of the relationship between the net H$\alpha$ emission
and the physical extent of the emitting circumstellar region produced
a clear correlation between the two properties for the first time.
The functional dependence between the parameters suggests that the
emission is directly proportional to the effective surface area of the
emitting disk.  These results are consistent with an optically thick
H$\alpha$ emission from a geometrically thin circumstellar disk.  We
do not detect any significant degree of correlation between the size
of the circumstellar region and the stellar rotation rate, although
the data suggest that hotter Be stars might have more extended
H$\alpha$-emitting regions.  Extending our analysis to a larger sample
of stars will verify if this is indeed a general property of classical
Be stars.

Because the formation and the subsequent evolution of the
circumstellar disks around Be stars are still the long-standing
puzzles associated with these stars, the interferometric
investigations by long-baseline interferometers will play a crucial
role in our understanding of these objects.  Already in the near-term
it should be possible to detect small-scale structures within the
circumstellar disks and investigate the characteristics of the inner
disk regions, thereby putting specific constraints on disk formation
theories.  Imaging small-scale structures, such as one-armed density
waves or spiral structures, will yield direct information about the
instabilities in the disks, possibly related in some cases to unseen
binary companions.  Also, combining spectroscopic and interferometric
observations, especially in the context of mass ejection
episodes~(which in some cases might be predicted based on
spectroscopy) will play a crucial role in our understanding of the
mass-feeding mechanisms, and the subsequent evolution of the material
in the disk.  Lastly, the direct detection of the deviation from
circular symmetry of the stellar surface, due to rapid rotation, will
help establish the connection between the stellar rotation and the
disk formation, independently of~(or in conjunction with)
spectroscopic studies.

The multi-spectral and the H$\alpha$-filtering capabilities of the
NPOI make it an optimal instrument to address the above described
issues.  We are currently setting up the NPOI for observations at
baselines of up to 100~m, which will allow us to study the spatial
properties of the H$\alpha$-emitting regions of Be stars in ways not
possible before.  Such studies have the potential of not only
contributing to our understanding of these interesting objects, but
they might also make a significant contribution to other branches of
stellar astrophysics.  For example, the investigation of disk
formation mechanism(s) will contribute to our understanding of
asymmetric mass-loss processes including stellar angular momentum
distribution~(especially in the context of rapidly rotating stars).
The apparent outflow of the material in the circumstellar disk and its
possible eventual dispersal, can be used as a test ground for
examining theories related to angular momentum transfer in outflowing
disks.

\bigskip
\bigskip

The Navy Prototype Optical Interferometer is a joint project of the
Naval Research Laboratory and the US Naval Observatory, in cooperation
with Lowell Observatory, and is funded by the Office of Naval Research
and the Oceanographer of the Navy.  We would like to thank Wes
Lockwood for accommodating our observing requests and for providing us
with the echelle spectra from the Lowell Observatory's Solar-Stellar
Spectrograph.  We thank David Mozurkewich and the anonymous referee
for useful comments on how to improve the manuscript.
C.T. acknowledges that this work was performed in part as a
Ph.D. research at the University of Toronto under funding from the
Government of Ontario Graduate Scholarship in Science and Technology
and Walter C.~Sumner Fellowship, as well as in part while being
employed by the NVI, Inc.~at the US Naval Observatory under contract
N0060001C0339.  Part of this work was also performed under a contract
with the Jet Propulsion Laboratory~(JPL) funded by NASA through the
Michelson Fellowship Program.  JPL is managed for NASA by the
California Institute of Technology.  This research has made use of the
SIMBAD literature database, operated at CDS, Strasbourg, France.

\newpage

%% The BIBLIOGRAPHY

%% TABLES %%

\newpage

\begin{table}[htp]
\caption[]{\sc \small Observing Log for $\eta$~Tauri and $\beta$~Canis~Minoris}
\begin{tabular}{lcc} \hline\hline
                              & $\eta$~Tau  & $\beta$~CMi \\
\hspace{2cm} Date\hspace{2cm} & \# of scans & \# of scans \\ \hline
   1997 Oct 15 \dotfill       &       5     &  \ldots \\
   1997 Nov 6  \dotfill       &      11     &     4   \\
   1997 Nov 7  \dotfill       &      10     &     2   \\
   1997 Nov 8  \dotfill       &   \ldots    &     3   \\
   1997 Nov 25 \dotfill       &       6     &  \ldots \\
   1997 Dec 5  \dotfill       &       3     &  \ldots \\
   1997 Dec 6  \dotfill       &       3     &  \ldots \\
   1998 Mar 22 \dotfill       &   \ldots    &     4   \\
   1998 Oct 7  \dotfill       &       5     &  \ldots \\
   1998 Oct 11 \dotfill       &       8     &  \ldots \\
   1998 Oct 12 \dotfill       &       3     &  \ldots \\
   1998 Nov 25 \dotfill       &       4     &     3   \\ 
   1998 Dec 11 \dotfill       &       2     &  \ldots \\
   1998 Dec 12 \dotfill       &       5     &  \ldots \\
   1998 Dec 24 \dotfill       &      18     &  \ldots \\
   1999 Feb 27 \dotfill       &       2     &     2   \\ 
   1999 Feb 28 \dotfill       &       3     &  \ldots \\ 
   1999 Mar 1  \dotfill       &       4     &  \ldots \\ 
   1999 Mar 4  \dotfill       &       2     &  \ldots \\ 
   1999 Mar 20 \dotfill       &       3     &  \ldots \\ 
   1999 Mar 25 \dotfill       &       3     &  \ldots \\ 
   1999 Apr 10 \dotfill       &  \ldots     &     2   \\
\hline
           Total              &     100     &  20 \\
\hline
\end{tabular}
\label{tab:new_observations}
\end{table}

\begin{table}[htp]
\caption[]{\sc \small Best-Fit Model Parameters for $\eta$~Tauri and $\beta$~Canis~Minoris}
\label{tab:new_parameters}
\begin{tabular}{cccccc} \hline\hline
            & $\theta_{\rm mj}$  &                   &    $\phi$   &                   &                \\ 
            &      (mas)         &        $r$        &    (deg)    &    $c_{\rm p}$    & $\chi^2_{\nu}$ \\ 
\hline 
$\eta$~Tau  &  $1.94 \pm 0.12$   &         1         &    $\ldots$ & $0.895 \pm 0.010$ & 1.03  \\ 
            &  $2.08 \pm 0.18$   & $0.753 \pm 0.053$ &$44.6\pm 9.4$& $0.893 \pm 0.015$ & 0.88  \\
\hline
$\beta$~CMi &  $2.02 \pm 0.28$   &         1         &    $\ldots$ & $0.904 \pm 0.019$ & 0.62  \\ 
            &  $2.13 \pm 0.50$   & $0.69 \pm 0.15$   & $40\pm 30$  & $0.874 \pm 0.069$ & 0.54  \\
\hline
\end{tabular}
\end{table}

\begin{table}[htp]
\caption{\sc \small The Circumstellar Regions of Be Stars}
\label{tab:be_star_sizes}
\begin{tabular}{rcccccccc} \hline\hline
       &              & Spectral &     $\pi$     &   $d$              & $\theta_{\rm mj}$&             &    $D_{\rm mj}$      \\
  HD   &     Star     & Type     &     (mas)     &   (pc)             &     (mas)        &   $r$       &    ($10^9$ m)    \\
\tiny{(1)} & \tiny{(2)} & \tiny{(3)} & \tiny{(4)} & \tiny{(5)}        & \tiny{(6)}       & \tiny{(7)}  &   \tiny{(8)}         \\   
\hline
5394   & $\gamma$ Cas & B0 IVe   & $5.32\pm0.56$ &$188^{+22}_{-18}$   &   3.67$\pm$0.09  & 0.79$\pm$0.03  & $103^{+13}_{-10}$    \\
23630  & $\eta$ Tau   & B7 IIIe  & $8.87\pm0.99$ &$113^{+14}_{-11}$   &   2.08$\pm$0.18  & 0.75$\pm$0.05  & $35.1^{+5.4}_{-4.7}$\\
37202  & $\zeta$ Tau  & B4 IIIpe & $7.82\pm1.02$ &$128^{+19}_{-15}$   &   3.14$\pm$0.21  & 0.31$\pm$0.07  & $60.1^{+9.9}_{-8.0}$\\
58715  & $\beta$ CMi  & B8 Ve    & $19.2\pm0.85$ &$52.2^{+2.4}_{-2.2}$&   2.13$\pm$0.50  & 0.69$\pm$0.15  & $16.6^{+4.0}_{-4.0}$ \\
\hline
10516  & $\phi$ Per   & B2 Vpe   & $4.55\pm0.75$ &$220^{+43}_{-31}$   &   2.67$\pm$0.20  & 0.46$\pm$0.04  & $87.8^{+19}_{-14}$  \\
22192  & $\psi$ Per   & B5 Ve    & $4.66\pm0.73$ &$215^{+40}_{-29}$   &   3.26$\pm$0.23  & 0.47$\pm$0.11  & $105^{+21}_{-16}$   \\
25940  & 48 Per       & B3 Ve    & $5.89\pm0.72$ &$170^{+24}_{-19}$   &   2.77$\pm$0.56  & 0.89$\pm$0.13  & $70.4^{+17}_{-16}$ \\
\hline
\end{tabular}\\[0.5ex]
\renewcommand{\baselinestretch}{1.0} \rm
\parbox{5.7in}{\footnotesize \quad {\sc Note.}--- Column~3:
spectral type from the BSC.  Column~4: parallax from the {\sl
Hipparcos} catalogue~\citep{ESA97}.  Column~5: distance to the source.
Column~6: the angular diameter of the major axis of the elliptical
Gaussian model representing the H$\alpha$-emitting envelope.
Column~7: axial ratio of the elliptical Gaussian model.  Column~8:
$\theta_{\rm mj}$ in physical units. }
\end{table}

\begin{table}[htp]
\caption[]{\sc \small Photometric Data for Be Stars}
\label{tab:be_star_mag}
\begin{tabular}{lcccccccc} \hline\hline 
\hspace{1cm} Star \hspace{1cm}    &  $V$  &  $B-V$  &$(B-V)_0$ &    $E(B-V)$   & $V-R$ & $E(V-R)$ &  $R_0$   \\
\hspace{1.3cm}\tiny{(1)}& \tiny{(2)} & \tiny{(3)} & \tiny{(4)} & \tiny{(5)} & \tiny{(6)} & \tiny{(7)} & \tiny{(8)} \\
\hline
$\gamma$ Cas \dotfill & 2.47  & $-0.15$ & $-0.30$  &   0.07        &  0.07 &  0.06    &   2.24   \\
$\eta$ Tau   \dotfill & 2.87  & $-0.09$ & $-0.12$  & $0.03^{\ast}$ &  0.03 &  0.02    &   2.77   \\
$\zeta$ Tau \dotfill  & 3.00  & $-0.19$ & $-0.18$  &   0.05        &$-0.03$&  0.04    &   2.91   \\              
$\beta$ CMi \dotfill  & 2.90  & $-0.09$ & $-0.11$  & $0.02^{\ast}$ &$-0.01$&  0.02    &   2.86   \\
\hline
$\phi$ Per  \dotfill  & 4.07  & $-0.04$ & $-0.24$  &   0.11        &  0.16 &  0.09    &   3.65   \\
$\psi$ Per  \dotfill  & 4.23  & $-0.06$ & $-0.16$  &   0.10        &  0.10 &  0.08    &   3.89   \\
48 Per      \dotfill  & 4.04  & $-0.03$ & $-0.20$  &   0.17        &  0.12 &  0.13    &   3.52   \\
\hline
\end{tabular}\\[0.5ex]
\renewcommand{\baselinestretch}{1.0} \rm
\parbox{5.8in}{\footnotesize \quad {\sc Note.}--- Columns~2 and 3: The
apparent $V$ magnitude and $B-V$ color from the BSC.  Column~4:
Intrinsic $(B-V)_0$ color based on the spectral
type~\citep{FitzGerald70}.  Column~5: $E(B-V)$ based on the absorption
bump at 2200~\AA~\citep{Beeckmans80}, except where denoted by
$^{\ast}$ for values obtained from the estimated $(B-V)_0$ color.
Column~6: $V-R$ colors from~\citet{Johnson66}.  Column~7: $E(V-R)$
derived from $E(V-R)/E(B-V) = 0.78\pm 0.02$~\citep{Schultz75}.
Column~8: Absolute $R$ magnitude corrected for interstellar
extinction.  }
\end{table}

\begin{table}[htp]
\caption{\sc \small Estimated Stellar Properties}
\label{tab:be_star_lum}
\begin{tabular}{lccccc} \hline\hline 
\hspace{1.5cm}          &   $f_{\star}$           & $4\pi d^2 f_{\star}$    &  $T_{\rm eff}$ & $R_{\star}$  &$4\pi R_{\star}^2 \mathcal{F}_{\star}^{\rm S}$\\
\hspace{0.3cm} Star   & ($10^{-12}$ W/m$^2$/nm) &      ($10^{25}$ W/nm)   & (K) $\pm$5\%   & ($R_{\sun}$) & ($10^{25}$ W/nm) \\
\hline
$\gamma$ Cas \dotfill & 2.71                    &  $114^{+27}_{-22}$      &    31,000      & 8$\pm$2      & $97\pm50$    \\
$\eta$ Tau   \dotfill & 1.65                    &  $25.2^{+6.3}_{-5.1}$   &    12,400      & 5$\pm$1      & $8.60\pm3.5$  \\
$\zeta$ Tau \dotfill  & 1.45                    &  $28.4^{+8.5}_{-6.5}$   &    16,400      & 6$\pm$1      & $18.8\pm6.6$  \\   
$\beta$ CMi \dotfill  & 1.52                    &  $4.95^{+0.46}_{-0.42}$ &    12,000      & 3.6$\pm$0.3  & $4.14\pm0.81$ \\
\hline
$\phi$ Per  \dotfill  & 0.735                   &  $42.5^{+17}_{-12}$     &    21,500      & 6.0$\pm$0.2  & $28.5\pm3.4$   \\
$\psi$ Per  \dotfill  & 0.587                   &  $32.4^{+12}_{-8.8}$    &    15,400      & 4.7$\pm$0.3  & $10.5\pm1.7$   \\
48 Per      \dotfill  & 0.830                   &  $28.6^{+8.0}_{-6.2}$   &    19,000      & 5.1$\pm$0.1  & $16.9\pm1.8$  \\
\hline
\end{tabular}\\[0.5ex]
\renewcommand{\baselinestretch}{1.0} \rm
\parbox{6.0in}{\footnotesize \quad {\sc Note.}--- $f_{\star}$ is the
physical flux (surface brightness) and $4\pi d^2 f_{\star}$~(where $d$
is the distance to the source from Table~\ref{tab:be_star_sizes}) is
the corresponding total radiation in the H$\alpha$ region based on
photometric data from Table~\ref{tab:be_star_mag}.  The effective
temperature~($T_{\rm eff}$) and the stellar radius~($R_{\star}$) are
based on the spectral type and tabulations of~\citet{Bohm-Vitense81}
and \citet{Underhill79}, respectively.  }
\end{table}

\begin{table}[htp]
\caption{\sc \small The H$\alpha$ Emission of Be Stars}
\label{tab:be_star_emission}
\begin{tabular}{lcccc} \hline\hline
    \hspace{3cm}     & $W_{{\rm H}\alpha}$ & $E_{{\rm H}\alpha}$ &   $L_{{\rm H}\alpha}$   &   $L_{{\rm H}\alpha}^{\rm S}$   \\
\hspace{1.0cm} Star  &     (nm)            &      (nm)           &   ($10^{25}$ W)     &   ($10^{25}$ W)             \\ 
\hline
$\gamma$ Cas\dotfill & $-2.25 \pm 0.11$  &  $-2.51 \pm 0.12$     &  $287^{+69}_{-56}$      &  $250\pm130$            \\
$\eta$ Tau  \dotfill & $-0.426\pm 0.021$ &  $-1.02 \pm 0.03$     &  $25.6^{+6.5}_{-5.2}$   &  $8.7\pm3.6$            \\
$\zeta$ Tau \dotfill & $-2.06 \pm 0.10$  &  $-2.55 \pm 0.11$     &  $72.2^{+22}_{-17}$     &  $48\pm17$              \\
$\beta$ CMi \dotfill & $-0.334\pm 0.017$ &  $-0.942\pm 0.026$    &  $4.66^{+0.45}_{-0.42}$ &  $3.90\pm0.77$          \\ 
\hline
$\phi$ Per  \dotfill & $-3.50 \pm 0.18$  &  $-3.87 \pm 0.18$     &  $165^{+65}_{-47}$      &  $110\pm14$               \\
$\psi$ Per  \dotfill & $-3.37 \pm 0.17$  &  $-3.89 \pm 0.17$     &  $126^{+47}_{-35}$      &  $40.8\pm6.9$                \\
48 Per      \dotfill & $-2.23 \pm 0.11$  &  $-2.68 \pm 0.12$     &  $77^{+22}_{-17}$       &  $45.3\pm5.2$                \\
\hline
\end{tabular}\\[0.5ex]
\renewcommand{\baselinestretch}{1.0} \rm
\parbox{5.5in}{\footnotesize \quad {\sc Note.}--- The H$\alpha$
equivalent widths~($W_{{\rm H}\alpha}$) are taken from a number of
different sources~(see \S~\ref{sec:data}) and their uncertainties are
assumed to be at 5\% level to account for observational errors as well
as possible intrinsic variability.  }
\end{table}

\begin{table}[htp]
\caption{\sc \small Rotational Velocities of Be Stars}
\label{tab:be_star_rotation}
\begin{tabular}{lcccc} \hline\hline
     \hspace{3cm}    &   $v\sin i$     &  $v_{\rm est}$  & $v_{\rm crit}$  &                            \\
\hspace{1.0cm} Star  & (km s$^{-1}$)   &  (km s$^{-1}$)  &   (km s$^{-1}$) & $v_{\rm est}/v_{\rm crit}$ \\
\hline
$\gamma$ Cas\dotfill &     230$\pm$23  &  375$\pm$40      &  538$\pm$54    &  0.70$\pm$0.10  \\
$\eta$ Tau  \dotfill &     140$\pm$14  &  212$\pm$28      &  406$\pm$41    &  0.53$\pm$0.09  \\
$\zeta$ Tau \dotfill &     220$\pm$22  &  231$\pm$28      &  441$\pm$44    &  0.52$\pm$0.08  \\
$\beta$ CMi \dotfill &     245$\pm$25  &  338$\pm$75      &  397$\pm$40    &  0.85$\pm$0.21  \\
\hline
$\phi$ Per  \dotfill &     400$\pm$40  &  450$\pm$46      &  477$\pm$48    &  0.94$\pm$0.14  \\
$\psi$ Per  \dotfill &     280$\pm$28  &  317$\pm$38      &  428$\pm$43    &  0.74$\pm$0.12  \\
48 Per      \dotfill &     200$\pm$20  &  $\gtrsim308$  &  458$\pm$46    &   $\gtrsim0.67$   \\
\hline
\end{tabular}\\[0.5ex]
\renewcommand{\baselinestretch}{1.0} \rm
\parbox{4.5in}{\footnotesize \quad {\sc Note.}--- The $v\sin i$ values
are from \citet{Slettebak82}, and $v_{\rm crit}$ values are from
\citet{Porter96}. }
\end{table}

%% FIGURES %%
\newpage

\end{document}